\documentclass[twocolumn,twocolappendix]{aastex631}
\usepackage[version=4]{mhchem}
\usepackage{upgreek}
\usepackage{multirow}

\received{March 26, 2021 }
\revised{September 2, 2021}
\accepted{October 8, 2021}
\submitjournal{The Planetary Science Journal}

\usepackage{xcolor}
\usepackage{soul}
\definecolor{yellowhl}{rgb}{0,1,1}
\definecolor{greyhl}{rgb}{0.8,0.8,0.8}
\definecolor{redhl}{rgb}{1,0.5,0.5}
\definecolor{bluehl}{rgb}{0.75,0.75,1}
\definecolor{greenhl}{rgb}{0.5,1.0,0.5}
\definecolor{purplehl}{rgb}{0.75,0.5,0.9}
\definecolor{orangehl}{rgb}{1,0.4,0.3}
\definecolor{pinkhl}{rgb}{0.9,0.4,0.9}

\graphicspath{{./}{figures/}}

\shorttitle{Proto-Mercury's Atmospheric Loss}
\shortauthors{Jaeggi et al.}

\begin{document}
\title{EVOLUTION OF MERCURY'S EARLIEST ATMOSPHERE}
\correspondingauthor{Noah J\"aggi}
\email{noah.jaeggi@space.unibe.ch}

\author[0000-0002-2740-7965]{Noah J\"aggi}
\affil{Physics Institute, University of Bern, Sidlerstrasse 5, 3012 Bern, Switzerland}
\author[0000-0002-7019-6286]{Diana Gamborino}
\affil{Physics Institute, University of Bern, Sidlerstrasse 5, 3012 Bern, Switzerland}
\author[0000-0002-0673-4860]{Dan J. Bower}
\affil{Center for Space and Habitability, University of Bern, Gesellschaftsstrasse 6, 3012 Bern, Switzerland}
\author[0000-0002-1462-1882]{Paolo A. Sossi}
\affil{Institute of Geochemistry and Petrology, Department of Earth Sciences, ETH Zurich, Clausiusstrasse 25, 8092 Zurich, Switzerland}
\author[0000-0003-2415-0508]{Aaron S. Wolf}
\affil{Earth and Environmental Sciences, University of Michigan, 1100 North University Avenue, Ann Arbor, MI 48109-1005, USA}
\author[0000-0002-1655-0715]{Apurva V. Oza}
\affil{Jet Propulsion Laboratory, California Institute of Technology, Pasadena, USA}
\affil{Physics Institute, University of Bern, Sidlerstrasse 5, 3012 Bern, Switzerland}
\author[0000-0002-7400-9142]{Audrey Vorburger}
\affil{Physics Institute, University of Bern, Sidlerstrasse 5, 3012 Bern, Switzerland}
\author[0000-0003-2425-3793]{Andr\'e Galli}
\affiliation{Physics Institute, University of Bern, Sidlerstrasse 5, 3012 Bern, Switzerland}
\author[0000-0002-2603-1169]{Peter Wurz}
\affil{Physics Institute, University of Bern, Sidlerstrasse 5, 3012 Bern, Switzerland}

\renewcommand{\edit}[2]{#2}


\begin{abstract}

\edit1{MESSENGER observations suggest a magma ocean formed on proto-Mercury, during which evaporation of metals and outgassing of C- and H-bearing volatiles produced an early atmosphere.  Atmospheric escape subsequently occurred by plasma heating, photoevaporation, Jeans escape, and photoionization.  To quantify atmospheric loss, we combine constraints on the lifetime of surficial melt, melt composition, and atmospheric composition.  Consideration of two initial Mercury sizes and four magma ocean compositions determine the atmospheric speciation at a given surface temperature.  A coupled interior--atmosphere model determines the cooling rate and therefore the lifetime of surficial melt. Combining the melt lifetime and escape flux calculations provide estimates for the total mass loss from early Mercury.  Loss rates by Jeans escape are negligible. Plasma heating and photoionization are limited by homopause diffusion rates of $\sim10^{6}$ kg/s.  Loss by photoevaporation depends on the timing of Mercury formation and assumed heating efficiency and ranges from $\sim10^{6.6}$ to $\sim10^{9.6}$ kg/s. The material for photoevaporation is sourced from below the homopause and is therefore energy-limited rather than diffusion-limited.  The timescale for efficient interior--atmosphere chemical exchange is less than ten thousand years. Therefore, escape processes only account for an equivalent loss of less than 2.3~km of crust ($0.3\%$ of Mercury's mass). Accordingly, $\leq0.02\%$ of the total mass of H$_2$O and Na is lost. Therefore, cumulative loss cannot significantly modify Mercury's bulk mantle composition during the magma ocean stage. Mercury's high core:mantle ratio and volatile-rich surface may instead reflect chemical variations in its building blocks resulting from its solar-proximal accretion environment.}

\end{abstract}

\section{Introduction} \label{intro}

MESSENGER data from X-ray, gamma ray, and neutron spectrometers constrain the composition of Mercury's surface, and motivate theories and models to understand Mercury's bulk composition, formation, and evolution.  The surface composition and geology of Mercury is compatible with partial melting of cumulates that were originally formed by magma ocean crystallization \citep{MPM18}.  Subsequent impact excavation exposed the cumulates at the surface \citep{MPM18,CGZ13}.  The low oxygen fugacity (\emph{f}\ce{O_2}) of the uppermost layer of Mercury's regolith, together with Mercury's large core size, suggest a reduced mantle where nominally lithophile elements such as Ca, Mn, Cr, and Ti are present in sulfides rather than silicates \citep{VKM16}.  Relative to basaltic rocks exposed at the surface of other terrestrial planets, a large amount of the moderately volatile element Na (3--5 wt\%) is detected on Mercury's surface \citep{Peplowski2014}. Observations of Na variation in Mercury's exosphere may relate to night-side deposit formation and dawn re-emission \citep[e.g.,][]{CMK16}. Hence, it remains an open question how moderately volatile elements such as Na may have accumulated on the surface---whether from an extant or now extinct process---and how their abundance compares to Mercury's bulk composition.  

Magma oceans are pivotal in determining the initial conditions and subsequent evolution and chemical differentiation of terrestrial planets in the solar system \citep[e.g.,][]{E12,Chao2021}. Radiometric dating reveals that magmatic iron meteorites, which represent planetesimal cores, formed within 2 Myr of solar system formation \citep{Kruijer2014}. The rocky planet whose mass is most similar to that of Mercury, and for which samples are available, Mars, likely accreted, formed an iron core, and underwent complete solidification of its magma ocean ocean within about 20 Myr of solar system formation \citep{BCC18}. Crucially, rapid core formation in terrestrial planets requires a magma ocean to enable efficient metal segregation \citep{S90}. By analogy, and given its solar-proximal location, extensive melting is therefore expected to have occurred on proto-Mercury \citep{VKM16, BE09}. Following its crystallisation, partial melting of magma ocean cumulates have been invoked to explain Mercury's contemporary surface composition \citep{MPM18}.

Energy from accretion and radiogenic heat (e.g. from $^{26}$Al) may have driven the differentiation of Mercury if it formed sufficiently early in solar system history \citep{BHS17,Siegfried1974}.  Following a phase of rapid growth, the subsequent reduction of impactor flux would have enabled Mercury's magma ocean to cool and crystallize without additional large-scale remelting. During this time, the mantle is expected to have stratified into a basal layer of olivine and a plagioclase and clinopyroxene dominated crust which is now observed on Mercury's surface \citep{BE09}. During the cooling of the magma ocean when the surface remains mostly molten, chemical species readily exchange between the interior, atmosphere, and exosphere---as occurred for other terrestrial planets in the solar system \citep[e.g.,][]{ET08}.

\cite{Feg87} addressed the hypothesis that the anomalously high bulk density of Mercury (owing to a high core/mantle ratio) is the result of evaporation of silicate melt components from the surface of a Hermean magma ocean. They presumed atmospheric loss was sufficiently slow that the atmosphere remained in equilibrium with the magma ocean.  In their model, vapor was removed in a step-wise fashion and the composition of the magma ocean evolved accordingly. In reality, however, evaporated species are transported, mixed, and lost from the atmosphere and exosphere, with the flux at which loss occurs integrated over the magma ocean lifetime ultimately dictating the total mass loss. Therefore, consideration of interior, atmospheric, and exospheric processes are necessary to assess whether significant quantities of rock-derived atmospheres can be lost during the Hermean magma ocean stage.

Based on observations of solar-mass stars, the early solar extreme ultraviolet (EUV) and X-ray fluxes were likely ~400 times larger than they are today.  This would have made photoionization a highly efficient non-thermal, and photoevaporation a highly efficient thermal, atmospheric escape mechanism \citep{Johnstone2015,TCG15}.  Other loss mechanisms of potential importance include atmospheric sputtering and kinetic escape (e.g. Jeans escape) that occur over the lifetime of the magma ocean.  Non-thermal loss rates can be constrained by known plasma pressures at proto-Mercury due to the incoming solar wind, as well as EUV luminosities of the early Sun estimated from population studies of nearby Sun-like stars \citep{ribas14,TCG15}.

In this paper we establish the extent of element evaporation and loss from Mercury during its early magma ocean phase. Models are constructed of the (1) coupled evolution of the magma ocean and atmosphere, (2) evaporation of metals and metal-oxide species from the Hermean magma ocean, (3) the mixing ratios and abundances of molecular species throughout the atmosphere and at the exobase, and finally (4) loss rates of these species from the upper atmosphere. We discuss these results in the context of the chemical evolution of Mercury's surface environment, \edit1{bulk composition,} and present-day observations.

\section{METHOD}\label{sec:methods}
\subsection{Overview}
Our combined modeling strategy provides insight into the initial composition and evolution of Mercury's exosphere by considering (1--2) energy and mass exchange \edit1{between the interior and atmosphere}, (3) speciation in the atmosphere, (4\edit1{-5}) loss from the exosphere:
\begin{enumerate}
 \item SPIDER \citep{BHS21,BKW19,BSW18} is a coupled interior--atmosphere model used to determine the \edit1{surface temperature and lifetime of melt at the surface}, as well as the pressure--temperature structure of the atmosphere. \edit1{Magma ocean cooling is regulated by the atmospheric opacity, which depends on the pressure of atmospheric species and hence the solubility of species in melt.}
 \item VapoRock \edit1{calculates} the equilibrium partial pressures of metal-bearing gas species of the elements Si, Al, Mg, Ca, Na, Fe and K above the magma ocean surface \citep[][]{Wolf2021}.  This determines the metal-bearing composition of the atmosphere as a function of temperature and the bulk composition of the magma ocean. It utilises ENKI's ThermoEngine (\url{http://enki-portal.org}) and combines estimates for element activities in silicate melts with thermodynamic data for metal and metal oxide vapor species \citep{Lamoreaux1987,Lamoreaux1984}. 
 \item VULCAN \citep{Tsai2017, Tsai2021} solves for the equilibrium chemistry of the atmosphere as a function of altitude by using element abundances for metals (output by VapoRock), volatile abundances (output by SPIDER), and the atmospheric pressure--temperature structure (also output from SPIDER). This provides the mixing ratios of atmospheric species, which are required to calculate escape at the exobase.
 \item DISHOOM \citep{OZA19,gebekoza2020} is an atmospheric evolution model which computes the total mass loss of \edit1{gaseous species} to space due to ultraviolet (EUV) heating, surface heating, as well as plasma heating.  
 \item Exospheric Monte Carlo (E-MC) model \citep{Gam19,VOR15,WULAM03} determines the rate of exospheric escape of particles due to Jeans escape and photoionization. It tracks particle trajectories using a thermal energy distribution that depends on the temperature at the exobase.
\end{enumerate}
\subsection{Cooling of the magma ocean} \label{sec:SPIDER}

Previous thermal modeling of Mercury's interior has focused either on the accretion phase \citep{BHS17} or its long-term evolution over billions of years \citep[e.g.,][]{TGP13,GBL11,S91,SSS83}.  Here, we model the thermal evolution of Mercury's magma ocean at the end of its accretion phase. At this time, the final magma ocean cools and crystallizes \edit1{on a timescale short enough such that there is negligible disruption} so our results remain independent of its accretion history. We model the thermal evolution of Mercury's magma ocean using SPIDER \citep{BSW18,BKW19} \edit1{to constrain the duration of melt at the surface as it cools from 2400 K to 1500 K.  This is necessary to compute the evaporation of metals and metal oxides at the planetary surface, prior to the formation of a surface lid around 1500 K}.  \edit1{Heating by the decay of radiogenic isotopes $^{26}$Al, $^{40}$K, $^{232}$Th, $^{235}$U, $^{238}$U is included in our model and the model starts from solar system time zero to obtain an upper estimate of the surface cooling time}.  The main parameters are provided in Table~\ref{tab:mopara} (Appendix~\ref{app:magmaocean})and are guided by the parameters and results from previous models of Mercury \citep{BHS17,TGP13}. 

Cases prefixed by `S' (`Small Mercury', Table~\ref{tab:mo_models}) have a planetary radius of 2440~km, which is the present-day radius of Mercury. Cases prefixed by `L' have a radius of 3290~km, which assumes Mercury was larger than at present day, perhaps due to mantle stripping driven by an impactor \citep{Chau2018,Asphaug2014,Benz2008}.  Cases with `V' (volatile) \edit1{consider the partitioning of carbon and hydrogen species, here termed volatiles, between the melt and atmosphere as well as redox reactions \citep{BHS21}}.  \edit1{By contrast, cases with `N' (non-volatile) do not consider volatiles but rather assume SiO is the only IR absorbing species; a suffix of `5' denotes a low SiO opacity ($10^{-5}$ m$^2$ kg$^{-1}$) and a suffix of `3' denotes a large SiO opacity ($10^{-3}$ m$^2$ kg$^{-1}$), both at a reference pressure of $3\times10^{-6}$ bar \citep[Fig.~2,][]{Semenov2003}.  For non-volatile cases, the surface pressure of SiO is imposed in SPIDER as a function of surface temperature as determined by VapoRock calculations.  Then the atmospheric opacity and hence magma ocean cooling rate can be determined.}

\begin{deluxetable}{@{\extracolsep{2pt}}llllllll@{}}
 \caption{
 Parameters for the magma ocean cases.
 \label{tab:mo_models}
 }
 \tablehead{
 \colhead{Case} & \colhead{$R_P$} & \colhead{$g$} &  \colhead{\ce{H_2O}} &  \colhead{\ce{H_2}} & \colhead{\ce{CO_2}} & \colhead{\ce{CO}} & \colhead{\ce{SiO}}\\
 \colhead{} & \colhead{km} & \colhead{m/s$^2$} & \multicolumn{5}{c}{Pressure (bar) at 2000 K}\\
 }
 \startdata
 SN5 & 2440 & 3.7  & ---  & --- & ---  & --- & $1.4E-4$ \\
 SN3 & 2440 & 3.7  & ---  & --- & ---  & --- & $1.4E-4$ \\
 SV & 2440 & 3.7  & 0.7  & 3.2 & 0.05 & 1.1 & --- \\
 LN5 & 3290 & 4.0  & ---  & --- & ---  & --- & $1.4E-4$ \\
 LN3 & 3290 & 4.0  & ---  & --- & ---  & --- & $1.4E-4$ \\
 LV & 3290 & 4.0  & 1.2  & 5.8 & 0.2  & 4.9 & --- \\
 \enddata
 \tablecomments{
 Small (S) and large (L) Mercury models are inspired by models M1 and M4 from \cite{BHS17}, respectively. Second letter of the case name denotes with volatiles (V) and no volatiles (N).  Non-volatile cases have an additional number of either 5 or 3, to denote small ($10^{-5}$ m$^2$ kg$^{-1}$) and large ($10^{-3}$ m$^2$ kg$^{-1}$) SiO opacity, respectively \citep{Semenov2003}.
 }
\end{deluxetable}


For cases SV and LV, carbon and hydrogen can either exist as reduced (CO, \ce{H_2}) or oxidised (\ce{CO_2}, \ce{H_2O}) species, where the \emph{f}\ce{O_2} is constrained to one log unit below the iron-wustite buffer ($\Delta$IW = $-1$, IW-1 hereafter). This is marginally higher than the most recent estimates for the \emph{f}\ce{O_2} of Mercury's mantle \citep{Cartier2019}.  For cases SV and LV, we determine the total H and C abundances by calculating the ppmw necessary for an Earth-size planet to produce a 100 bar \ce{CO_2} (i.e. Venus-like atmosphere) and 270 bar \ce{H_2O} (i.e. one Earth ocean mass) atmosphere.  \edit1{The abundances of H and C are equivalent to 330 ppmw of \ce{H_2O} and 120 ppmw of \ce{CO_2}, respectively.  The mass of large Mercury's mantle is about a factor of 5 larger than small Mercury's mantle, resulting in a 5 times increase in the total volatile budget.}

\subsection{Evaporation from the magma ocean} \label{sec:evap}
At the high surface temperatures that characterize a magma ocean ($>$1500~K), the partial pressures of the vapor species of the major rock-forming oxides (e.g. Si\ce{O_2}, NaO$_{0.5}$, KO$_{0.5}$) can become significant \citep{Sossi2018,Sossi2019,Visscher2013}. Gas-liquid equilibria for these elements are described by congruent evaporation, generalised as:

\begin{equation}
 \mathrm{M}^{x+n}\mathrm{O}_{(x+n)/2} (l) = \mathrm{M}^{x}\mathrm{O}_{x/2} (g) + \frac{n}{4} \mathrm{\ce{O_2}} (g), 
\label{eq:evap1} 
\end{equation}

\noindent where M is the metal, $x$ is the oxidation state of the metal in its gaseous state, and $n$ the number of electrons exchanged in the reaction. Both $x$  and $n$  are integer values that may be $\geq{0}$ or $\leq{0}$. At equilibrium, the partial pressure of any given metal or metal-oxide species in an ideal gas is given by:

\begin{equation}
 p(\mathrm{M}^{x}\mathrm{O}_{x/2}) = \frac{K_{(1)}X(\mathrm{M}^{x+n}\mathrm{O}_{(x+n)/2})\gamma(\mathrm{M}^{x+n}\mathrm{O}_{(x+n)/2})}{(f\mathrm{\ce{O_2}})^{(n/4)}},
\label{eq:evap2} 
\end{equation}

\noindent where $K_{(1)}$ is the equilibrium constant of the reaction (Eq.~\ref{eq:evap1}), $X$ is the mole fraction, and $\gamma$  is the activity coefficient of the metal oxide melt species, $\mathrm{M}^{x+n}\mathrm{O}_{(x+n)/2}$. Equilibrium constants involving 31 gas species (Table~\ref{tab:oxides}, Appendix~\ref{app:oxides}) are calculated according to their thermodynamic properties given in \cite{Lamoreaux1987,Lamoreaux1984}. Evident from Eq.~\ref{eq:evap2} is that estimates for the composition of the silicate melt in addition to the activity coefficients of its constituent components are required to correctly predict partial pressures.  To this end, likely compositions representative of Mercury's crust, mantle and possible precursors are shown in Table \ref{tab:mo_comp}. The MELTS algorithm is used to estimate activity coefficients of melt oxide species \citep{Ghiorso1995}. The \emph{f}\ce{O_2} is constrained to lie one log unit below the IW buffer (IW-1), which is calculated according to \cite{Oneill2002}:

\begin{equation}
\log f\mathrm{\ce{O_2}}(\mathrm{IW}) = 2\frac{-244118+115.559\,T-8.474\,T\,\mathrm{ln}(T)}{\mathrm{ln}(10)R\,T},
\label{eq:IW} 
\end{equation}

\noindent where $T$ is temperature and $R$ the gas constant. These ingredients together comprise the VapoRock code and permit calculation of equilibrium partial pressures over a range of temperatures, \emph{f}\ce{O_2}, and silicate melt compositions \citep[][]{Wolf2021}.
\subsection{Magma ocean composition} \label{sec:mo_comp}
The composition of the early Hermean mantle is uncertain. To address how variability in the surface composition affects the evolved partial pressures of metal and metal oxide gas species, four compositions are investigated (Table \ref{tab:mo_comp}): (1) Enstatite Chondrites (EH4), (2) Bencubbin Chondrites (CB), (3) Northern Smooth Plain (NSP) lava, and (4) NSP source. The first composition assumes that core--mantle differentiation was sluggish, with the composition of the magma ocean being approximated by Enstatite Chondrites (EH4), often cited as appropriate starting compositions for Mercury due to their high bulk iron content, strongly reduced nature, and the resemblance of partial melts thereof to Hermean surface compositions \citep{Nittler2011}. We assume that all FeO is extracted in the form of metallic iron to form Mercury's core, which results in high \ce{SiO2} and MgO contents in the complementary silicate fraction.

The second composition is based on chondrules found in the Bencubbin-class of Carbonaceous Chondrites (CB), which best reproduce Mercury's surface composition based on MESSENGER data \citep{MPM18,BE09}. Spectrometric measurements of Mercury's surface show a crust rich in Na and S and poor in Fe relative to other basaltic rocks \citep{Nittler2019}. Presuming these abundances are representative of bulk Mercury, two other compositions are investigated; the Northern Smooth Plain lava (NSP melt) that represents a volatile-rich composition observed on the surface, and its inferred mantle source (NSP source) \citep{Nittler2019}. These two compositions represent Mercury's crust and mantle, respectively. Although it is not anticipated that magma ocean crystallisation produced the NSP melt composition directly, it is included to define an end-member opposing the CB composition that is Na and K poor, and comparably Fe-rich (Table \ref{tab:mo_comp}).

\begin{deluxetable}{lDDDD}[!htbp]
\caption{Magma ocean surface composition.
\label{tab:mo_comp}}
\tablehead{
\colhead{\edit1{Oxide}} & \twocolhead{EH4}	& \twocolhead{CB}	& \twocolhead{NSP} & \twocolhead{NSP} \\
\colhead{\edit1{(wt\%)}}& \twocolhead{}   & \twocolhead{}    & \twocolhead{source} & \twocolhead{lava}
 }
\decimals
\startdata
SiO$_2$     & 62.73   & 50.70     &  53.67     &  58.70 \\
Al$_2$O$_3$ & 2.58    & 4.60      &  4.75      &  13.80 \\
MgO         & 30.24   & 36.90     &  36.89     &  13.90 \\
CaO         & 1.99    & 3.30      &   2.26     &  5.81 \\
FeO         & 0.00    & 3.50      &   0.02     &  0.04 \\
Na$_2$O     & 1.71    & 0.19      &   1.97     &  7.00 \\
K$_2$O      & 0.20    & 0.05      &   0.05     &  0.20 \\
Total       & 99.45   & 99.24     &  99.61     &  99.45
\enddata
\tablecomments{Compositions based on Enstatite chondrites (EH4) \citep{Wiik1956}; CB chondrite chondrule data with bulk CB Na and K mass balanced for chondrules to fit bulk meteorite iron-silicate ratio \citep{Lauretta2007,Weisberg2000,Weisberg1990}; and northern smooth plains (NSP) composition for the lava and source \citep{Nittler2019,Namur2016}}
\end{deluxetable}

\subsection{Atmospheric structure} \label{sec:atm}
The atmospheric structure is constrained by the temperature at the magma ocean--atmosphere interface, the planetary equilibrium temperature, and the atmospheric composition and pressure (Appendix~\ref{app:PTprofile}).  For\edit1{ non-volatile} cases, the calculated vapor pressures \edit1{of Si, Na, K, Fe, Mg, Al, and Ca oxide species} in equilibrium with the magma ocean (Fig.~\ref{fig:partvapP}) are used directly in the exospheric loss model (Section~\ref{sec:ionization}). \edit1{SiO vapour pressures reported in Table~\ref{tab:mo_models} at $T_\mathrm{surf}=2000$~K are not strongly affected by the magma ocean composition and range from $10^{-4\pm0.1}$ bar.} For cases SV and LV in which Mercury's atmosphere contains outgassed H- and C-bearing gases, the partial pressures of \ce{H_2}, \ce{H_2O}, CO, and \ce{CO_2} are calculated by SPIDER according to volatile solubility and \emph{f}\ce{O_2} buffered by the magma ocean at IW-1. A modified version of the VULCAN code is then used to compute the equilibrium chemical speciation in the atmosphere that contains both the metal-bearing gases and H and C volatiles \citep{Tsai2017,Tsai2021}. VULCAN computes the atmospheric mixing ratios using the pressure--temperature ($P$--$T$) structure of the atmosphere. 

VULCAN by default includes about 300 reactions for C, H, O, and N to which we added reactions involving Si, Mg, Ca, Fe, Na and K to obtain their equilibrium speciation (Table~\ref{tab:vulcreac}, Appendix~\ref{app:VULCAN}). Table~\ref{tab:vulcinit} shows the initial element-to-hydrogen ratios used in the VULCAN calculations.  \edit1{Surface vapor} pressures of Ca and Al-bearing species did not exceed $10^{-6}$ bar in the magma ocean temperature range investigated, and are thus excluded. For the remaining species we used a case dependent $P$--$T$ profile from SPIDER at a surface temperature of 2000~K to determine the mixing ratio of species in the atmosphere as a function of altitude. The $P$--$T$ profile may imply condensation of certain elements initially present in the vapor, \edit1{which may rain out of the atmosphere prior to escape}. To assess this possibility, Gibbs Free Energy minimisation of the atmospheric composition  were preformed \edit1{throughout the atmospheric column} using FactSage 7.3. \citep{Bale2016}. 


\begin{deluxetable*}{@{\extracolsep{4pt}}lccccccc@{}}[!htbp]
\tablewidth{0pt} 
\tablecaption{Surface element ratios at $T_\mathrm{surf} = 2000$~K.} \label{tab:vulcinit}
\tablehead{
\colhead{Composition} & \multicolumn7c{\edit1{Element Ratios}} \\
\cline{1-1}\cline{2-8}
{} & \colhead{C/H}  & \colhead{O/H} & \colhead{Mg/H} & \colhead{Si/H} & \colhead{Na/H} & \colhead{K/H} & \colhead{Fe/H}
} 
\startdata 
\cline{1-8}
\multicolumn{8}{c}{\edit1{SV Case}}\\
\cline{1-8}
EH4      			& \multirow{4}{*}{1.532E-01}    & \multirow{4}{*}{2.472E-01}& 1.251E-06 & 1.469E-05 & 1.254E-04 & 1.121E-05 & 1.053E-08 \\
CB           		&                               &                           & 1.680E-06 & 1.039E-05 & 7.155E-05 & 5.785E-06 & 9.388E-06 \\
NSP source          &                               &                           & 1.740E-06 & 1.020E-05 & 2.282E-04 & 5.254E-06 & 7.155E-07 \\
NSP melt        	&                               &                           & 9.642E-07 & 1.161E-05 & 3.480E-04 & 7.139E-06 & 1.351E-06 \\
\cline{1-8}
\multicolumn{8}{c}{\edit1{LV Case}}\\
\cline{1-8}
EH4                 & \multirow{4}{*}{3.608E-01}    & \multirow{4}{*}{4.641E-01}& 6.806E-07 & 7.997E-06 & 6.825E-05 & 6.104E-06 & 5.732E-09 \\
CB           		&                               &                           & 9.144E-07 & 5.653E-06 & 3.894E-05 & 3.148E-06 & 5.109E-06 \\
NSP source          &                               &                           & 9.469E-07 & 5.551E-06 & 1.242E-04 & 2.859E-06 & 3.894E-07 \\
NSP melt        	&                               &                           & 5.248E-07 & 6.321E-06 & 1.894E-04 & 3.886E-06 & 7.354E-07 \\
\enddata
\tablecomments{Ratios are based on SPIDER and VapoRock results for magma ocean compositions given in Table \ref{tab:mo_comp}. We neglect Al and Ca as their vapor pressure do not exceed $10^{-9}$ bar at 2000~K for any composition.}
\end{deluxetable*}

SPIDER, and \edit1{for volatile cases also} VULCAN, provide descriptions of the atmospheric structure and composition needed to determine the altitude of the \edit1{homopause and} exobase. \edit1{The homopause is the altitude at which molecular diffusion exceeds diffusion by eddies and thus separates the well-mixed lower atmosphere from the mass-separated upper atmosphere.} The exobase is the altitude at which gas is loosely bound to the planet and is collisionless (Knudsen number~=~1) resulting in efficient escape.  

\subsubsection{Homopause \edit1{level and} diffusion} \label{sec:homodiff}
\edit1{To determine the homopause level $z_\mathrm{hom}$ we require the particle density at the homopause $n_\mathrm{hom}$. For a steady-state homopause height, the molecular coefficient $D_{ik}$  is equal to the eddy diffusion coefficient $K_{zz}$, allowing us to solve for the particle density $n_\mathrm{hom}$.} The diffusion coefficient $D_{ik}$ (m$^2$/s) 
\edit1{within the homosphere} is calculated for each major species using the Chapman-Enskog relation \citep{Chapman1970}. It determines the binary diffusion rate of a gaseous species $i$ with mass $m_i$ within a gas of average mass $m_k$:

\begin{equation}
 D_{ik} = \frac{3 \sqrt{k_B T_\mathrm{skin}}}{8 n_\mathrm{hom}  \sigma^2_{ik}\Omega_{ik}}\sqrt{\frac{1}{2\pi}\frac{m_i+m_k}{m_i m_k}},
\label{eq:moldiff} 
\end{equation}

\noindent where, $k_B$ is the Boltzmann constant and $T_\mathrm{skin}$ \edit1{the absolute temperature at the homopause (skin temperature)}. \edit1{Homopause pressures ($P_\mathrm{hom}= n_\mathrm{hom}k_B T_\mathrm{skin}$) for each species are thereby about $10^{-6}$~bar for all cases.} We approximate the intermolecular distance $\sigma_{ik}=\frac{\sigma_i+ \sigma_k}{2}$ with the radius of the species relative to the mean species diameter weighted by the mixing ratio. The dimensionless collision integral $\Omega_{ik}$ is assumed to be unity. 



The eddy velocity is often approximated by the atmospheric species thermal speed $v_\mathrm{th}$, and the characteristic eddy length scale $L_{eddy}$ is approximated by the atmospheric scale height $H$ \citep[e.g.][]{Atreya1986}. Values for $K(z)= v_\mathrm{eddy}\,L_\mathrm{eddy}$  that are calculated based on this assumption exceed the suggested eddy diffusion coefficient upper limit of 320~m$^2$/s by several orders of magnitude \citep{VlaK15}. Hence we use this upper limit in the volatile cases to determine $n_{\mathrm{homo}}$, which is based on the energy dissipation rate within the Earth's atmosphere.

To compute the $z_\mathrm{hom}$ and $T_\mathrm{hom}$ for the volatile cases 
we determine the altitude at which the \edit1{$P$--$T$ profile reaches} a number density $n_{\mathrm{homo}}(K(z))$. As the number density at the homopause only depends on $K_{zz}$ with the same order of magnitude for both volatile and non-volatile cases, $n_{\mathrm{homo}}(K(z))$ is approximately $10^{18}-10^{19}\,\mathrm{at}\,\mathrm{m}^{-3}$. \edit1{ For the non-volatile cases we do not obtain a P(z)-T profile from VULCAN for the non-volatile cases due to the absence of H based species.   We therefore use the barometric formula with gravity as a function of height to compute $z_\mathrm{hom}$. }

\subsubsection{\edit1{Exobase level}}
\edit1{Due to the large difference in number density between the homopause and the exobase, the barometric formula is not applicable assuming an isothermal upper atmosphere with height-dependent gravity. We approximate the exobase height of early Mercury, which is subject to extensive loss, by finding an exobase height that results in a loss rate that is in equilibrium with the homopause diffusion rate. The loss from the exobase is proportional to the exobase height (increasing surface area) whereas diffusion from the homopause to the exobase is inversely proportional to the exobase height (decreasing density gradient). The exobase altitude $z_\mathrm{exo}$ of each species is determined for all cases by setting the homopause diffusion rate $\dot{M}_{\mathrm{diff},i}$ equal to the largest, diffusion limited mass loss rate of photoionization $\dot{M}_\mathrm{ion}$ (Eq.~\ref{eq:photoionization}, Section~\ref{sec:ionization})}.\\

The homopause diffusion rate $\dot{M}_\mathrm{diff}$ in kg/s of a species $i$ is obtained by multiplying the diffusion coefficient $D_{ik}$ with the species number density gradient, the species mixing ratio at the homopause $n_{i}/n_{hom}$, and the homopause surface area $A_\mathrm{hom}$: 

\begin{equation}
\dot{M}_{\mathrm{diff},i} = -D_{ik} \frac{\Delta n_\mathrm{exo-hom}}{\Delta z_\mathrm{exo-hom}}\,m_i \frac{n_{i}}{n_\mathrm{hom}}\,A_\mathrm{hom}.
\label{eq:diffusionrate}
\end{equation}

The number density of particles at the exobase $n_{\mathrm{exo}}$ is necessary to determine \edit1{the number density gradient between the homopause and exobase and ultimately} $z_\mathrm{exo}$. \edit1{For a single species $i$}, the exobase is defined at a altitude at which the particle free path ($\lambda_{col}=1/n_{\mathrm{exo}}\sigma_{\mathrm{col}}$) is equal to the exospheric scale height ($H = k_B T /m g$), therefore \citep[i.e.,][]{Gronoff2020}:

\begin{equation}
   n_{i,\mathrm{exo}} = \frac{1}{H_i\sigma_i} = \frac{m_i\,g(h)}{k_B\,T_\mathrm{skin}\,\sigma_i},
\label{eq:nexo}
\end{equation}

\noindent with the collision cross section $\sigma_i$, the skin temperature $T_\mathrm{skin}$, and the acceleration of gravity $g(h)$ at the exobase altitude $z_\mathrm{exo}$. In a multi-species atmosphere, each species has a specific mass and collision cross section, leading to a species-specific scale height and exobase density and altitude. The collision cross sections (CCSs) of each species are approximated as their respective atom or molecule size (Table \ref{tab:crosssec}, Appendix \ref{app:crosssections}). \edit1{Typical values for $n_{\mathrm{exo}}$ are around $10^{12}-10^{13}\,\mathrm{atom}\,\mathrm{m}^{-3}$ which coincides with $\sim10^{-12}$~bar}. 
The \edit1{skin temperature} used for determining $z_\mathrm{exo}$ and $z_\mathrm{hom}$ is derived from the atmosphere model (Appendix~\ref{app:PTprofile}):
\begin{equation}
T_\mathrm{skin} =  \left[  \frac{ \epsilon (T_\mathrm{surf}^4-T_\infty^4)} {2  }+T_\infty^4 \right]^{1/4} ,
\label{eq:Tskin}
\end{equation}
with the magma ocean surface temperature $T_\mathrm{surf}$, emissivity $\epsilon$ (depends on optical depth and hence atmospheric composition and pressure), and equilibrium temperature $T_\infty$.

\subsection{Exospheric loss} \label{sec:ionization}

The E-MC escape model focuses on Jeans escape, and photoionization and photodissociation to investigate the loss of proto-Mercury's exosphere. These mechanisms compete for importance; Jeans escape acts at high exospheric temperatures, whereas photoionization and photodissociation act at large solar EUV and X-ray fluxes present during early times, respectively. The E-MC model simulates escape by tracking $\sim10^5$ exospheric particles, with trajectories initiated at the exobase with a random angle and energy selected from a Maxwellian velocity distribution function \citep{VOR15}. The initial energy of the exospheric particles depends on the exobase temperature, which decreases with time due to magma ocean and atmospheric cooling.
 
The loss processes in the E-MC model are calculated on a particle by particle basis. As soon as a particle reaches Mercury's Hill radius (Table~\ref{tab:escape_parameters}), it is assumed to have escaped Mercury's gravitational attraction and is subsequently removed from the simulation. Another loss process is through interaction with photons. At each altitude step starting from the exobase and moving away from the planet, the E-MC model calculates the probability of a particle being photo-dissociated or photoionized. If the particle is photodissociated, the code calculates the corresponding trajectories of the fragments and assesses the chance of escaping the gravitational well and the potential for subsequent photoionization. Ionized particles are considered lost from the exosphere, \edit1{assuming} they are picked up by the electromagnetic forces of the solar wind plasma or Mercury's magnetospheric plasma. Photoionization and photodissociation rates are scaled for each dominant species using the EUV flux of the early Sun. The EUV flux and mass loss are dependent on the rotational evolution of the Sun, with a fast rotator being much more active than a slow rotator \citep{Johnstone2015, TCG15}. We consider a moderately fast rotating Sun, where the EUV luminosity $L_{\mathrm{EUV}}$ (J/s) is:

\begin{equation}
    L_{\mathrm{EUV}} = (4.7\times10^{25})\ t^{-1.18},
\label{eq:lumin}
\end{equation}

\noindent where $t$ (Ma) is the time since the formation of the solar system. \edit1{Typical values for the incident EUV fluxes at Mercury at 1~Ma and 5~Ma are thereby $10^{3.0}$ and $10^{2.2}$ J/s m$^{-2}$ respectively.}

The loss is equal to the sum of the exospheric particles that have either been photoionized or lost through gravitational escape. The loss of a given species from the exosphere at a given time is then calculated using its mixing ratio at the exobase. For a given exobase temperature, the loss rate \edit1{($\dot{M}=dm/dt$)} from the exosphere by photoionization $\dot{M}_\mathrm{ion}$ is:

\begin{equation}
\dot{M}_\mathrm{ion} = \sum_{i=0}\,\psi^\mathrm{source}_i\,\xi_\mathrm{i}\, m_i A_\mathrm{exo},
\label{eq:photoionization}    
\end{equation}

\noindent where $A_{\mathrm{exo}}$ is the surface area of the exobase and $0 \leq \xi_i \leq 1$ is the fraction of lost particles of a species $i$ with mass $m_i$, thermal speed of  $ v_i^{\mathrm{therm}}$, and particle flux leaving the exobase of $\psi^{\mathrm{source}}_{\mathrm{i}}=n^{exo}_i\, v_i^{therm}$. The area of the exobase is equal to sum of the $R_P$ and $z_\mathrm{exo}$ (Section~\ref{sec:atm}). The total loss is determined by integrating the loss flux over the lifetime of \edit1{surficial melt}. 

\subsection{Atmospheric loss and surface evaporation}\label{sec:heating}
Atmospheric loss by thermal processes (photoevaporation) and non-thermal processes (plasma-heating) are determined using DISHOOM \citep{OZA19}, using Eq.~\ref{eq:photoevap} and Eq.~\ref{eq:plasmaloss}, respectively. Preliminary calculation of Jeans-escape using DISHOOM demonstrated negligible loss due to surface heating compared to all other mechanisms. The escape parameters appropriate to proto-Mercury are summarized in Table~\ref{tab:escape_parameters}. 

\begin{deluxetable}{@{\extracolsep{4pt}}cccccc@{}}[!htbp]
\label{tab:escape_parameters}
\tablewidth{0pt} 
\tablecaption{
Escape parameters $\lambda_{0}$ for small Mercury and large Mercury cases at $T_\mathrm{surf}=2000$~K.
}
\tablehead{\colhead{Species} & \colhead{$\lambda_{0}$}& \colhead{$\mu_\mathrm{atm}$} & \colhead{$T_\mathrm{exo}$}   &  \colhead{$z_\mathrm{exo}$}  & \colhead{Case}\\
  \colhead{}    & \colhead{}&\colhead{[amu]} & \colhead{[K]}  &  \colhead{[km]} & \colhead{}}
\startdata
    \cline{1-6}\\[-1ex]
    \multicolumn6c{Small Mercury, $R_H$~=~72 $R_P$}\\
    \cline{1-6}                                             
       \edit1{Na, K, Fe}   & \edit1{7.8}  & 24.1           & \edit1{1613}     & \edit1{2670} & SN5, SN3\\
       H, C, O              & \edit1{7.6}  & \edit1{14.3}  & \edit1{1021}     & \edit1{2370} & SV\\[2ex]
    \multicolumn6c{Large Mercury, $R_H$~=~90 $R_P$}\\
    \cline{1-6}                                             
       \edit1{Na, K, Fe}   & \edit1{15.7} & 24.1           & \edit1{1615}             & \edit1{1890} & LN5, LN3\\
       H, C, O              & \edit1{15.3} & 14.3           & \edit1{893}              & \edit1{2160} & LV\\
\enddata
\tablecomments{The escape parameter $\lambda_{0}$ with the respective mean molecular weight of the upper atmosphere $\mu_\mathrm{atm}$, as well as the exobase temperature $T_\mathrm{exo}$ and altitude $z_\mathrm{exo}$. The Hill radius $R_H$ in Mercury radii $R_P$ describes the gravitational field of influence of Mercury in each case.}
\end{deluxetable}

Irradiation from the impinging solar wind plasma and high energy photons may heat the atmosphere and drive escape at a level that is significantly larger than surface heating and photoionization. Upper-atmospheric heating (photoevaporation) is caused by incoming X-ray and EUV photons that deposit heat into a neutral medium via molecular absorption \citep{WAT81} or photoelectric heating \citep{Murray-Clay2009}.  This expands the atmospheric envelope beyond the gravitational influence of the body ($R_H$ in Table~\ref{tab:escape_parameters}). The heating can be estimated by energy-limited escape driven by EUV photons \citep{WAT81}, \edit1{which is a reasonable approximation to thermally-driven hydrodynamic escape \citep{Krenn2021,Volkov2013}}:

\begin{equation}
\begin{split}
\dot{M}_U & = \frac{\eta_\mathrm{EUV}\,L_\mathrm{EUV}\,2\pi\,z_\mathrm{abs}^2 }{U_\mathrm{env}},
\end{split}
\label{eq:photoevap}    
\end{equation}

\noindent where, $z_\mathrm{abs}$ is the absorption altitude, generally taken to be 1.25 planetary radii for an outgassing atmosphere \citep[e.g.,][]{Johnson2015} where the X-ray and EUV photons can absorb and thereby deposit heat into the atmospheric molecules. We note that we use 1.25 Mercury radii $R_P$, as a conservative lower-limit in absorption altitude as the homopause situated at $\sim$ 1.4 $R_P$ represents an upper-limit.  The efficiency at which the atmosphere is heated, $\eta_{EUV}$, is uncertain so \edit1{we use $10^{-3}$ as a conservative lower estimate \citep{Ito2021} and $10^{-1}$ as an upper limit \citep{Mordasini2020}.} \edit1{Both efficiencies that we used were previously applied to atmospheres that use vastly different planet parameters but similar enough as they consider a metal oxide (non-volatile) or a H/He (volatile) atmosphere respectively.} Hot Jupiter H/He envelopes as well as volcanic atmospheres suggest $\eta_{EUV}$ may be as large as $0.35$ \citep{Murray-Clay2009, Lellouch1992}. 

Mass loss due to plasma-heating is observed at Jupiter's moon Io and is fundamentally driven by plasma ram pressure and magnetic pressure interacting with the atmosphere \citep[e.g.,][]{johnson1990}. Therefore, we estimate the atmospheric loss from an impinging plasma on Mercury by scaling to the plasma pressures measured at the Galilean satellites \citep{Johnson2004}.  Following \cite{OZA19,gebekoza2020}, the mass loss of a species $i$ by plasma heating at proto-Mercury is:

\begin{equation}
\begin{split}
\dot{M}_{P, i} & = \frac{x_i}{\dot{M}_{i,\mathrm{Io}}} \, \hat{P} \, \hat{U} \, \hat{v}_\mathrm{ion} \, \left(\hat{h}_\mathrm{exo}\right)^2
\end{split}
\label{eq:plasmaloss}    
\end{equation}

\noindent where $x_i$ is the element fraction of the species $i$ in the atmosphere, and $\dot{M}_{i,\mathrm{Io}}$ its atmospheric sputtering loss rate at Io. The total plasma pressure $\hat{P}$, gravitational binding energy $\hat{U}$, and ion velocity $\hat{v}_\mathrm{ion}$ of Mercury are expressed as non-dimensional values that are scaled to Io's corresponding values. The total plasma pressure is additive where $P_\mathrm{tot} = P_\mathrm{mag}+P_\mathrm{ram}$. For the calculations we use a magnetic pressure $P_{mag} = 1.7$ nPa based on an estimation of Mercury's magnetic moment of 2.76 $\times10^{12}$~T~m$^{3}$ at the magnetopause stand-off distance of 1.4 $R_P$.  The ram pressure due to the solar wind varies from $\sim$ 10--30 nPa \citep{Korth2012}, yielding a total pressure of $\sim$ 12--32 nPa. 

\edit1{Evaporation from the magma ocean and atmospheric loss have to be equal to retain the atmospheric pressure and thus a steady state. The evaporation rate of a species is approximated by the Hertz-Knudsen-Langmuir equation. The evaporation rate of a species $i$ with molar mass $M_i$ over the surface of Mercury in mol/s is given by:}

\begin{equation}
\dot{M}_{\mathrm{evap},i} = M_i \frac{dn_i}{dt} 
= -4\pi R_P^2M_i\frac{\gamma_{ev}P_{i,eq}-\gamma_{en}P_{i,s}}{\sqrt{2\pi M_iRT}},
\label{eq:diff_surf}
\end{equation}
\noindent \edit1{with the evaporation and condensation coefficients $\gamma$ (set to unity for a liquid), surface pressure $P_s$, equilibrium pressure $P_{eq}$, and the Mercury radius $R_P$. By setting the homopause diffusion rate $\dot{M}_{\mathrm{diff},i}$ equal to the evaporation rate $\dot{M}_{\mathrm{evap},i}$, the equation is solved for the ratio of surface to equilibrium pressure, $p_{i,s}/p_{i,eq}$ for each species at each temperature step. For homopause diffusion rates, the ratio lies $>$~0.99, therefore the atmosphere up to the homopause is considered to be in equilibrium.}

\section{RESULTS}\label{sec:results}
\subsection{Surficial melt lifetime and atmospheric structure}\label{sec:res_cooling}
\begin{figure}[tbhp]
\centering
\plotone{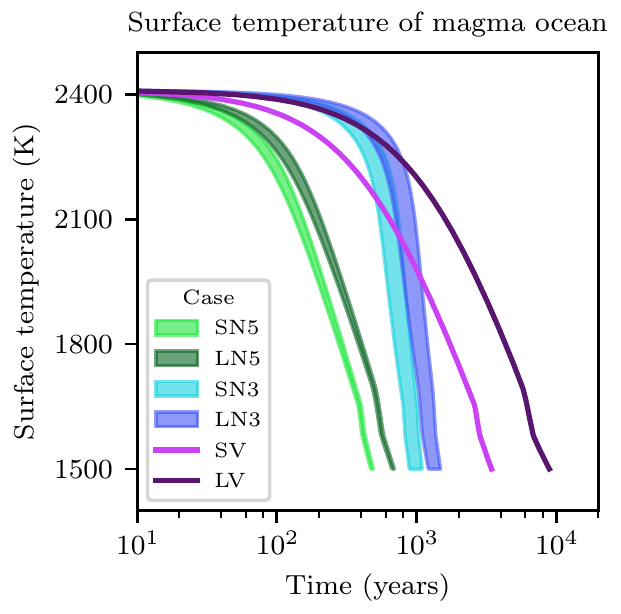}
\caption{Evolution of the surface temperature of the Hermean magma ocean.  \edit1{For non-volatile cases, the upper bound of cooling is provided by EH4 composition and the lower bound by Cb composition}.  See Table~\ref{tab:mo_models} for case parameters.}
\label{fig:mo_cooling}
\end{figure}

The surface temperature of the Hermean magma ocean cools from 2400 to 1500~K in around 400 -- 9000~years, depending on the planetary size (i.e. mantle mass) and efficiency of radiative energy loss to space (Fig.~\ref{fig:mo_cooling}).  The cooling rate is inversely proportional to $R_P$ since it depends on the ratio of the planetary surface area to mantle mass.  Hence, a large Mercury takes longer to cool than a small Mercury for otherwise identical parameters.

\subsubsection{Non-volatile cases}
\edit1{The cooling trajectory of non-volatile cases is characterised by two episodes.  First, when the surface temperature is high (early time), the pressure and hence opacity of SiO is large and therefore cooling is slow.  The second episode of cooling is rapid, since even a small decrease in surface temperature produces a drastic fall in both SiO pressure and opacity, driving the planet towards cooling like an ideal black body.}  Therefore, the cooling timescale for N5 cases is only marginally greater than for an ideal black body, which bounds the minimum cooling time to \edit1{470}~years. \edit1{Increasing SiO opacity by 2 orders of magnitude} increases the \edit1{minimum cooling time to around 1000~years (N3 cases), marginally affected by the chosen magma ocean composition.}

\begin{figure*}[!ht]
\gridline{\fig{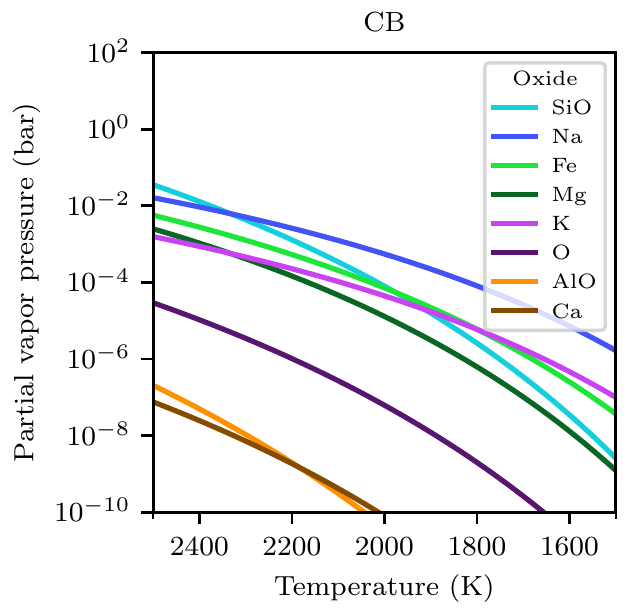}{0.45\textwidth}{}
        \fig{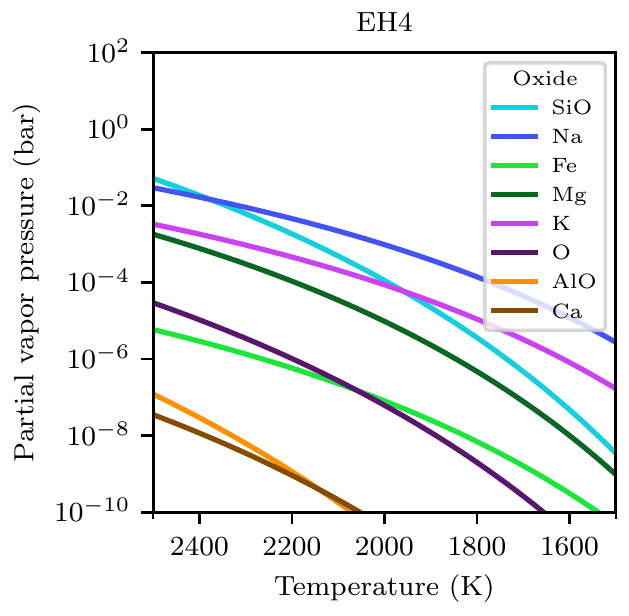}{0.45\textwidth}{}}
\gridline{\fig{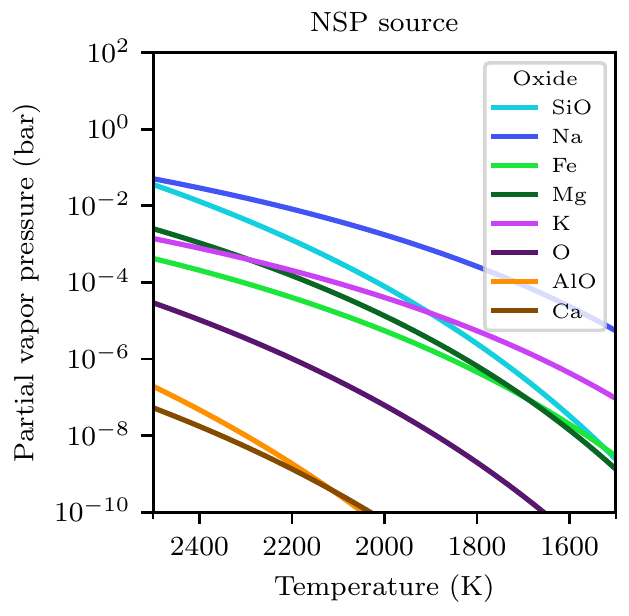}{0.45\textwidth}{}
         \fig{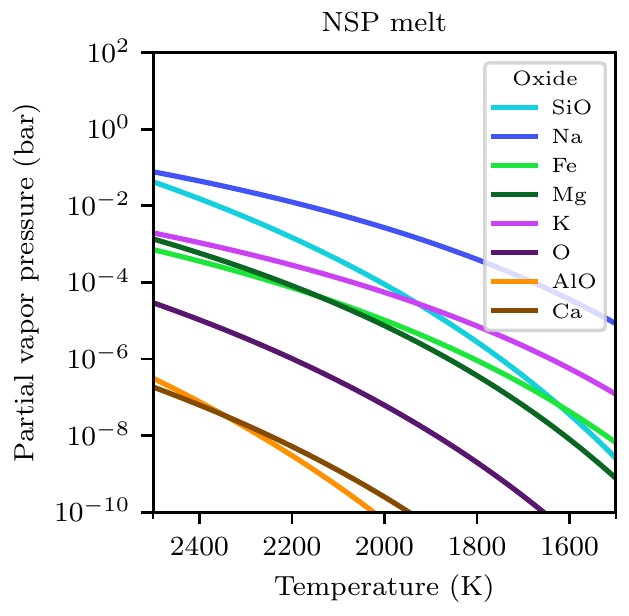}{0.45\textwidth}{}
         }
\caption{
Oxide partial vapor pressures  calculated  using  the  CB, EH4 and NSP source and NSP melt composition. The  mixing ratios are thereby reflected by the activities of the elements in the given melt composition. The high-Fe composition CB shows a large Fe Partial pressure, comparable with K in the high-K compositions EH4 and NSP melt.
}
\label{fig:partvapP}
\end{figure*}

Non-volatile cases at 2400~K have thin atmospheres of $< 0.1$~bar and comprise $>99.8\%$ gaseous Na, SiO, Fe, K, and Mg. The major constituents are Na and SiO at high temperatures whereas the mixing ratio of SiO rapidly decreases below 2400~K (Fig. \ref{fig:partvapP}). The partial pressure of Mg behaves similar to SiO but does not exceed $1\%$ of the atmospheric mixing ratio for any composition. The mixing ratios of Fe and K, however, are more variable because they depend on the assumed magma ocean composition and reach their highest mixing ratios of \edit1{ $9\%$ and $2\%$, respectively, for CB and EH4} compositions with high FeO and \ce{K2O} (Table \ref{tab:mo_comp}, Fig.~\ref{fig:partvapP}). Refractory components---AlO and Ca---have negligible partial pressures ($<10^{-6}$ bar at 2400~K) and are thus ignored in further calculations. The highest total surface pressure of metal-bearing species at low temperatures is obtained with the NSP melt composition. At 2400~K, the total surface pressure is $6.16\times10^{-2}$~bar, which decreases to $8.52\times10^{-6}$~bar at 1500~K. 

\begin{figure*} [htb!]
\gridline{\fig{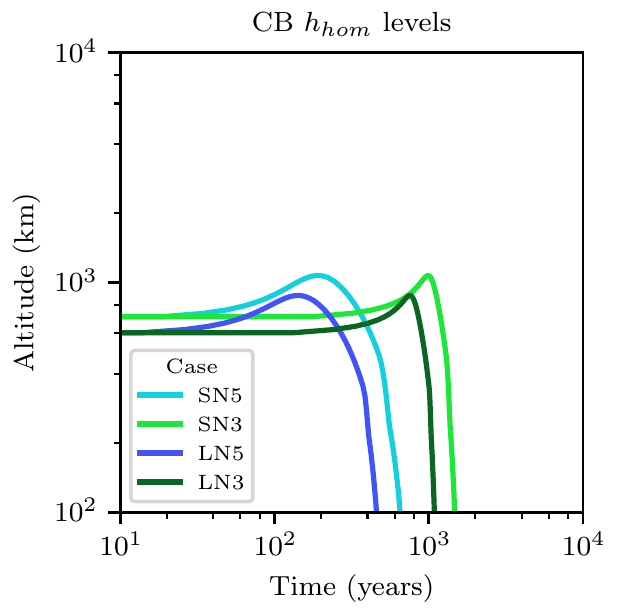}{0.45\textwidth}{}
          \fig{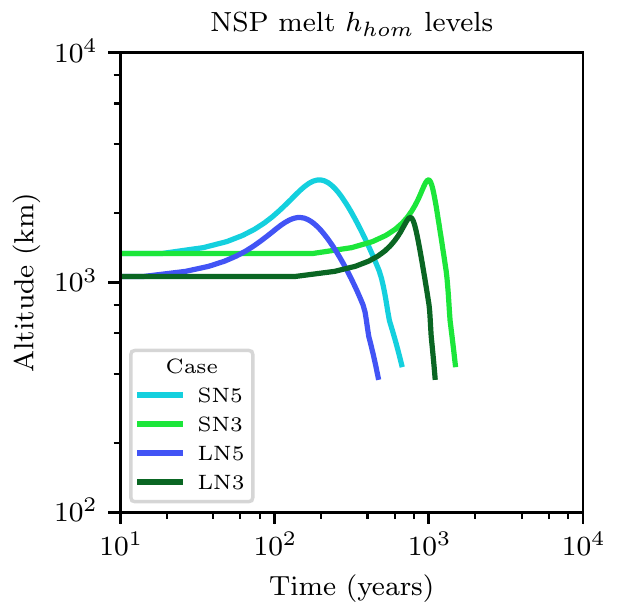}{0.45\textwidth}{}
         }
\caption{\edit1{Homopause altitudes for sodium} derived for the \edit1{non-volatile cooling scenarios }for CB and NSP melt. Lines terminate when the surface temperature reaches 1500~K. \edit1{The equilibrium exobase levels are not plotted as they lie close to the homopause levels. Homopause levels for other species show the same trends and reach comparable elevations.}}
\label{fig:hom_levels}
\end{figure*}

Figure~\ref{fig:hom_levels} shows the homopause levels \edit1{of Na} for the non-volatile cases as a function of time for CB and NSP melt compositions. \edit1{Exobase levels lie within a few 100s of kilometers of the homopause and are omitted in the log-log plot as a result}. During the magma ocean phase, the levels evolve within \edit1{470--660~years for N5 cases, and 1100--1480~years for N3} cases. The homopause and exobase locations are only weakly sensitive to the planet size and gravity. The magma ocean composition, however, exerts a strong influence on the atmospheric structure. For the CB case, the homopause lies at \edit1{685~km whereas for the NSP melt composition it lies around 1258~km at a magma ocean surface temperature of 2400~K.} \edit2{The early inflation of an atmosphere above a cooling magma ocean is due to increasing $T_\mathrm{skin}$ caused by decreasing IR-opacity as the partial pressure of SiO decreases (Eq. \ref{eq:Tskin}). Following this stage, the homopause altitude falls to 83~km (CB) and 439~km (NSP melt) at 1500~K.} The exobase density and location is further dependent on the mean cross section of atmospheric species, which is tied to the composition-dependent vapor pressures (Eq. \ref{eq:nexo}). The high vapor pressure of Na in the NSP melt relative to the CB composition lowers the mean molecular weight and the mean collision cross section of the atmosphere, which both increase its extent.  

\subsubsection{Volatile cases}
Volatile-bearing cases result in cooling times of \edit1{3400}~years (Case SV) and \edit1{8900}~years (Case LV). Both small and large proto-Mercury have the same initial volatile abundances of C and H by ppmw, but this manifests in a larger total reservoir size of volatiles for a large proto-Mercury compared to a small one. The mass of volatiles in the atmosphere defines the surface atmospheric pressure, which in turn determines the optical thickness of the atmosphere and hence the efficiency of radiative cooling.

\begin{figure*}
    \centering
    \includegraphics{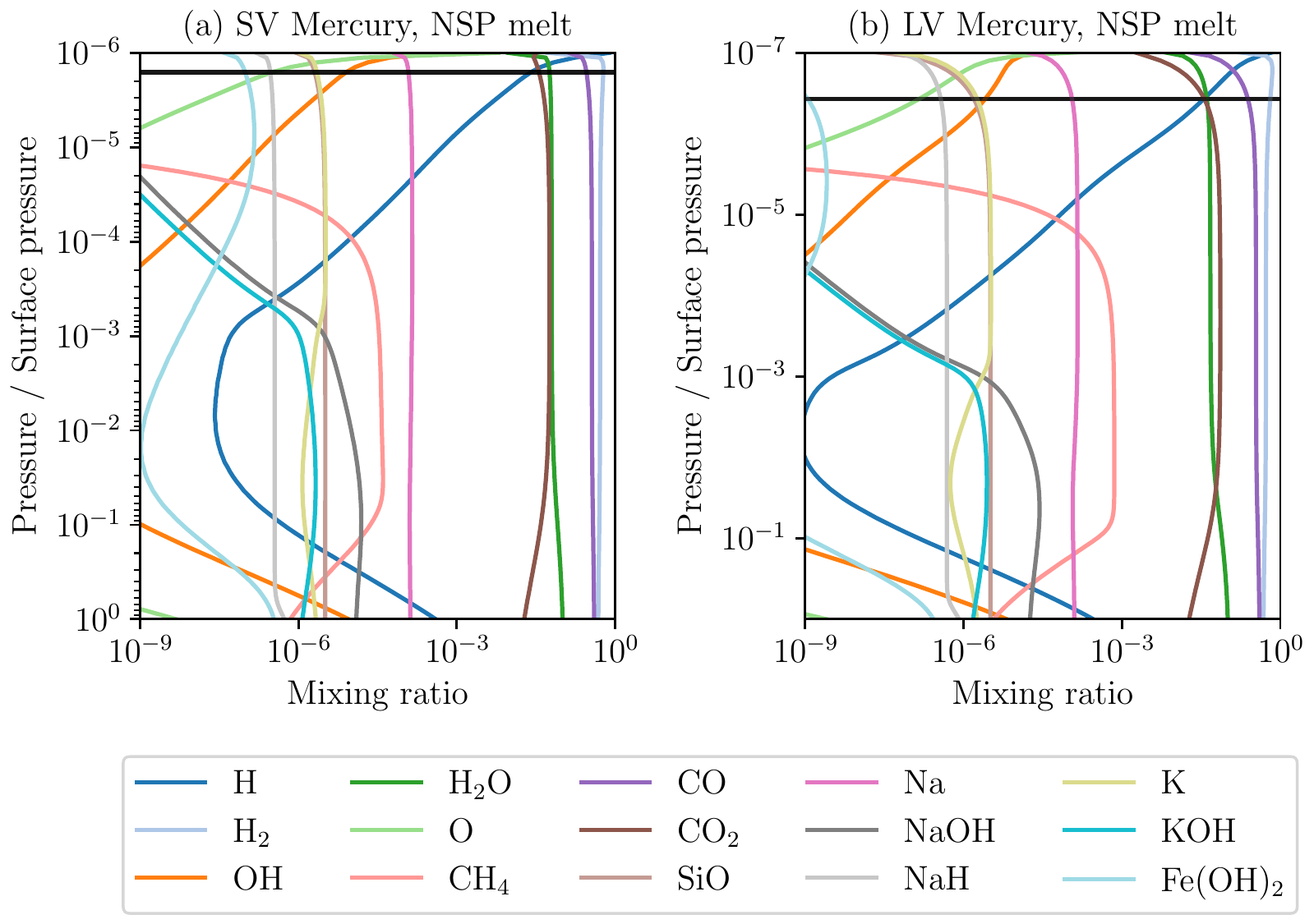}
    \caption{Major and minor element composition of the atmosphere for NSP melt composition with a surface temperature of 2000~K for (a) SV and (b) LV Mercury models.  The surface pressure that normalises the y-axis is \edit1{5.0}~bar for SV and \edit1{12.1}~bar for LV. \edit1{The homopause level of H is plotted as a black, horizontal line.}
    }
    \label{fig:VULCANresults}
\end{figure*}

Atmospheres of volatile cases around a small and large proto-Mercury reach surface pressures of about 5 and 12 bar at a magma ocean surface temperature $T_\mathrm{surf}$ of 2000~K, respectively (Fig.~\ref{fig:VULCANresults}). This result is independent of the partial pressures of the metal-bearing species, as their contribution is $\leq0.1$ bar at $T_\mathrm{surf}$~=~2000~K. Thus, it is the outgassed hydrogen and carbon species (which depends on their solubilities) that dictates the surface pressure.  VULCAN is then used to compute the equilibrium chemistry of the atmosphere accounting for the outgassed volatiles as well as the metals and oxides. \edit1{For both small and large Mercury, the atmosphere below the homopause is dominated} by \ce{H2} and CO with about \edit1{60 and 27 vol.}\%, respectively, at $T_\mathrm{surf}$~=~2000~K. 
\edit1{Between the homopause and the exobase} the dominant \edit1{H, C and O based species dissociate to monoatomic gases}.

For all compositions, Na and K are the dominant metallic elements at $T_\mathrm{surf}=2000$~K. At the surface \edit1{their hydroxide forms NaOH and KOH are fairly abundant, but dissociate towards the homopause (Table~\ref{fig:VULCANresults})}  
\edit1{Sodium hydride (NaH) is also present at the surface at pressures one order of magnitude lower than NaOH and remains about constant throughout the atmosphere, reaching similar mixing ratios to K.  Potassium hydride (KH) is ignored as no rate constant exists in the NIST kinetics database (\url{kinetics.nist.gov}).}

Exobase levels are situated \edit1{up to 2910 and 2590~km for small and large Mercury with homopause levels down to 2360 and 2160}~km, respectively. The $P$--$T$ structure of the atmosphere gives $T_\mathrm{skin}$ values for the homopause and exobase of about \edit1{1021 and 893~K} (Table \ref{tab:escape_parameters}), respectively, for small and large cases, which lies well below the $T_\mathrm{skin}$ calculated for the non-volatile cases ($T_\mathrm{skin}=1615$~K). The exobase levels of the volatile cases are thus \edit1{comparable to} the non-volatile cases. 
Unlike the non-volatile cases, the planet size has a large impact on the atmospheric structure, which is solely due to the difference in the total volatile reservoir (Section~\ref{sec:res_cooling}). 


\subsection{Atmospheric loss}
We find that the major atmospheric escape mechanism is photoevaporation (Eq.~\ref{eq:photoevap}) with a lower limit of photoevaporation constrained to \edit1{10$^{6.6}$~kg/s }and an upper-limit of \edit1{10$^{9.6}$~kg/s} for both non-volatile (Na) and volatile cases (H, C, and O). \edit1{The upper limit is thereby} roughly three orders of magnitude larger than the photoionization of the major atmospheric species. \edit1{For the high heating efficiency ($\eta_\mathrm{EUV}=10^{-1}$) case, loss rates become evaporation limited when reaching 1600~K as the surface to equilibrium pressure ratio approaches zero. For low heating efficiencies ($\eta_\mathrm{EUV}=10^{-3}$), the ratio of surface to equilibrium pressure $p_{i,s}/p_{i,eq}$  remains at $>$0.93 (Eq.~\ref{eq:diff_surf}). Photoevaporation as an approximation of thermally-driven hydrodynamic escape (Eq.~\ref{eq:photoevap}) therefore expresses the highest uncertainty on the stability of the atmosphere.}

\begin{figure}[tbhp]
\centering
\plotone{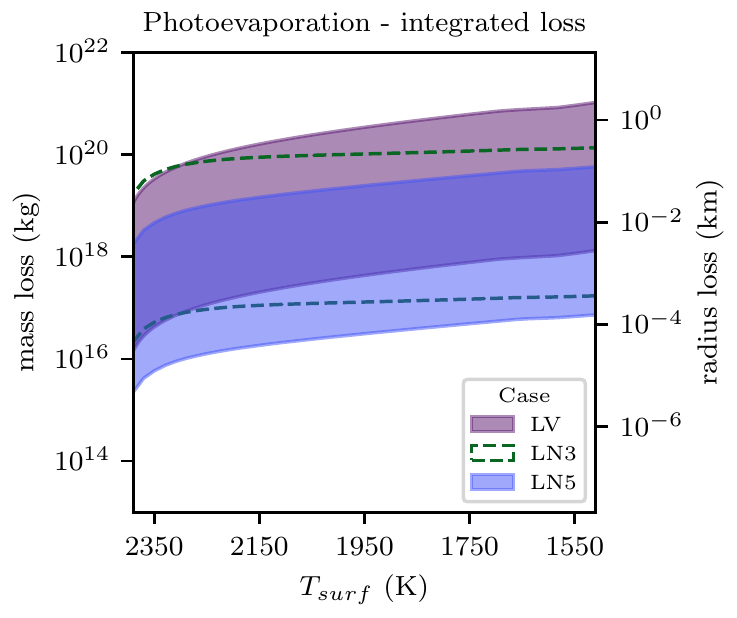}
\caption{Integrated mass loss rates of the bulk atmosphere through photoevaporation from 2400~K to 1500~K for large Mercury cases. \edit1{The shaded areas represent the uncertainty of loss rates based on initial conditions. For the N3 case only the upper and lower limit are shown as dashed lines as it lies within both V and N5 areas. The highest loss rates assume upper limits for EUV luminosity only 1~Ma after Sun formation, L$_\mathrm{EUV}$(1~Ma), and heating efficiency of $10^{-1}$, and the lowest assume L$_\mathrm{EUV}$(5~Ma), and heating efficiency of $10^{-3}$. Using a mantle density of $\rho_\mathrm{mantle}$~=~3.5~g/cm$^3$ results in a maximum of crust material being lost to space ranging between 2.3~km and 16~cm, depending on the degree of XUV intensity and heating efficiency.}\label{fig:photoevap_integrated}}
\end{figure}

Figure~\ref{fig:photoevap_integrated} shows the integrated mass loss over the most extensive \edit1{surficial melt lifetime} of \edit1{8900~years}. The photoevaporative erosion $dR$ of the surface can be estimated by assuming mass conservation where:

\begin{equation}
    \frac{dM}{dR} = 4 \pi R_P^2 \rho_{mantle}.
\end{equation}

Using a mantle density of $\rho_{mantle}$~=~3.5~g/cm$^3$ 
\edit1{and} assuming \edit1{a high EUV heating efficiency of $10^{-1}$ at a large Mercury size} allows for \edit1{$\sim$ 2.3~km loss of crust over the 8900~years} of the volatile case \edit1{surficial melt} lifetime. \edit1{We have shown using Eq.~\ref{eq:diff_surf} that this case becomes evaporation limited due to the high photoevaporation rates. The total integrated loss when assuming an EUV heating efficiency of $10^{-1}$ is therefore lower than shown in Figure~\ref{fig:photoevap_integrated} but not significantly. This is due to most loss occurring during the early magma ocean stage, when the surface temperatures are high and evaporation is not limiting the potentially high photoevaporation rates.}

Plasma-driven escape is diffusion-limited as a supply is required at the exobase (Eq.~\ref{eq:diffusionrate}). The $\dot{M}_P$ calculated \edit1{lie orders of magnitudes below} $\dot{M}_\mathrm{ion}$ as given in (Table \ref{tab:lossrates}) \edit1{and therefore do not affect the atmospheric structure}. Using an intermediate total pressure of 25 nPa, we find about $\dot{M}_P$~=~10$^{3.4}$~kg/s for small and large, non-volatile Mercury cases and about $\dot{M}_P$~=~10$^{3.1}$~kg/s for volatile Mercury cases, respectively.  The loss rates are thereby comparable to the plasma-driven escape observed on Jupiter's moon Io in \ce{SO_2}  $\sim$ 10$^{3}$~kg/s \citep[][]{thomas2004}. 

Time-averaged mass loss rates by photoionization are given in (Table \ref{tab:lossrates}). Like non-thermal plasma-driven escape, non-thermal escape due to photoionization at the exobase (Eq.~\ref{eq:photoionization}) is diffusion-limited. 
As \edit1{photoionization} rates of non-volatile cases follow the same trends independent of planet size, we report results focusing on a small proto-Mercury only, omitting the large Mercury \edit1{photoionization} rates, which are mostly \edit1{within a factor two for the dominant species (Table \ref{tab:lossrates})}. \edit1{The results of non-volatile N5 cases are also not reported, as an almost isothermal atmosphere results in less than a factor two larger loss rates at high temperatures. }

\edit1{Regarding the presence of metal oxide derived gaseous species,} Gibbs Free Energy minimisation of the vapor phase (using FactSage) along a case-dependent atmospheric $P$--$T$ profile (Figure \ref{fig:PTprofile}, Appendix \ref{app:PTprofile}) indicates that Mg and SiO condense into clinopyroxene (1900~K) and then into olivine ($\approx$ 1700~K) during cooling, by which temperature their fraction remaining in the gas is negligible. Iron persists in the vapor to lower temperatures, condensing partially into olivine before iron metal condenses at 1350~K. Therefore, while Mg, Si (and Ca and Al) all condense prior to reaching the exobase ($T \approx$ 1680~K), Fe is likely to partially reside in the vapor phase. Sodium never fully condenses (Nepheline, its major host mineral, condenses in very minor proportions below 1500~K), while K remains entirely in the vapor phase down to at least 950~K. 

\subsubsection{Loss from non-volatile atmospheres} \label{sec:nonvol_loss}
\begin{figure*} [htb!]
\gridline{
          \fig{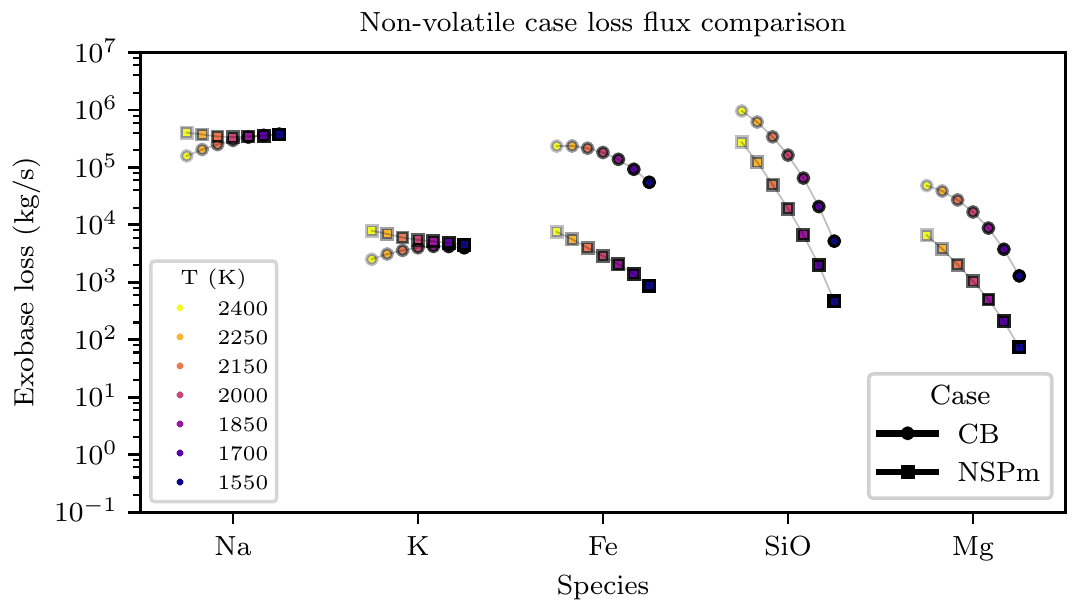}{0.90\textwidth}{}
        }
\caption{
Mass loss fluxes of exospheric species from 2400~K to \edit1{1550}~K plotted for the \edit2{small Mercury} \edit1{N3} cases. The loss fluxes for large Mercury follow the same trends and are about a factor of two \edit2{smaller}. \edit1{The N5 cases with isothermal atmospheres show a factor of two larger loss rates at high temperatures but the same species related trends. }
\label{fig:BGexoflux}}
\end{figure*}

$\dot{M}_\mathrm{ion}$ in non-volatile cases is sensitive to the chosen initial composition (Fig.~\ref{fig:BGexoflux}). The loss fluxes of the non-volatile species of interest ---SiO, Na and K---are proportional to their mixing ratios in the atmosphere (Fig.~\ref{fig:partvapP}). \edit1{In cases with high initial \ce{SiO} partial pressures} the mixing ratios and hence the diffusion-limited loss rates of Na and K increase during initial cooling as SiO becomes less abundant. This is most evident in the CB loss flux with an \edit1{initially} increasing loss rate despite decreasing temperatures (Fig.~\ref{fig:BGexoflux}). 

\begin{figure*}[tbhp]
\gridline{\fig{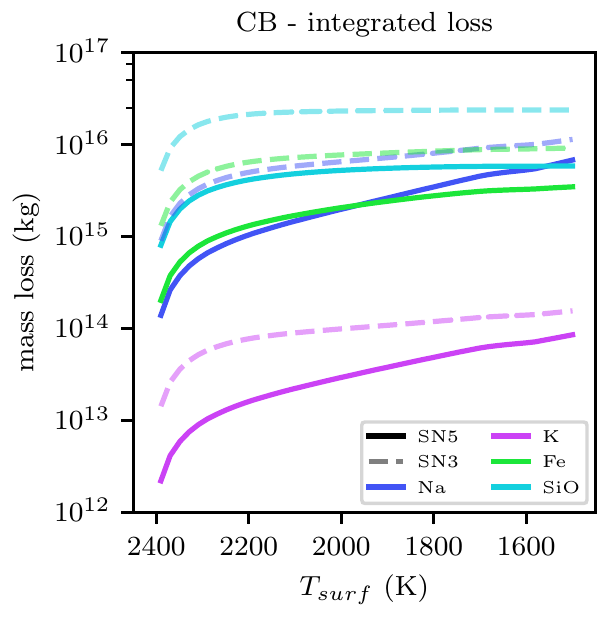}{0.45\textwidth}{}
    \fig{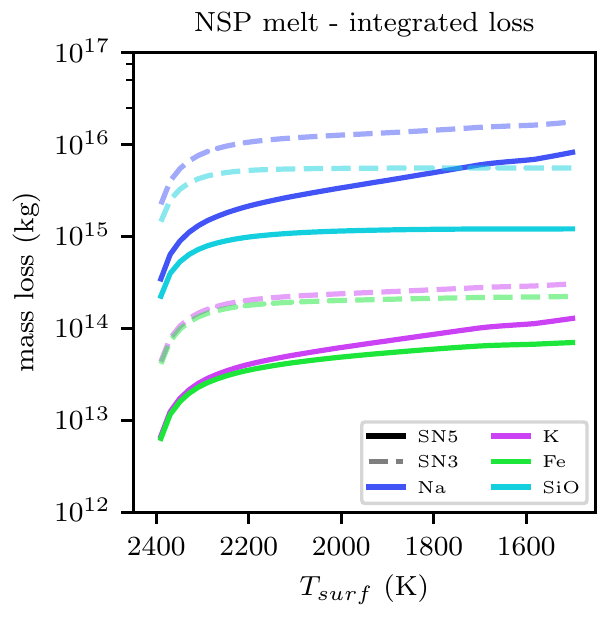}{0.45\textwidth}{}}
\caption{Integrated mass loss rates of exospheric oxides and elements through photoionization as the magma ocean cools from a surface temperature of 2400~K to 1500~K. \edit1{SiO loss is shown although it is unlikely to persist in the upper atmosphere as it is consistently below its highest condensation temperature of $T=1900$.}
\label{fig:BGintegrate}}
\end{figure*}

In the low-Na and low-K composition CB, where SiO is the dominant metal oxide at high temperatures, loss fluxes of SiO reach up to \edit1{$9.6\times10^5$~kg/s at $T_\mathrm{surface}$~=~2400~K. The vapor pressure of SiO declines with respect to other dominant gas species (Na and K), thereby reducing its mixing ratio rapidly with decreasing temperature. The lower mixing ratio of SiO causes loss rates to drop to $9.4\times10^2$~kg/s at 1500~K. For the same composition, loss rates for Na and K are around $2.0\times10^5$~kg/s and $3.1\times10^3$~kg/s at $T_\mathrm{surf}$~=~2400~K, increasing to $3.2\times10^5$~kg/s and $4.2\times10^3$~kg/s at $T_\mathrm{surf}$~=~1500~K, respectively.} For the high-Na end-member composition of NSP melt, the diffusion-limited loss rates for Na and K are $5.9\times10^5$~kg/s and $1.2\times10^4$~kg/s respectively when at $T_\mathrm{surf}$~=~2400~K, and decrease to $3.9\times10^5$~kg/s and $4.3\times10^3$~kg/s respectively when $T_\mathrm{surf}$~=~1500~K. The ratio of Na to K diffusion rates is $\sim$ 100 and hence about one order of magnitude higher than their mixing ratios in the atmosphere. Integrated diffusion-limited losses over the \edit1{surficial melt lifetimes} are shown in Figure~\ref{fig:BGintegrate}.


If we assume \edit1{based on FactSage results}, that SiO, and Mg are absent and only Na, K and Fe remain in the atmosphere, the diffusion-limited loss rates for all cases peak in the \edit2{small Mercury NSPm case with a fairly temperature independent Na loss rate of about $10^6$~kg/s.} \edit1{Loss rates for K and Fe are thereby about three orders of magnitude lower than that of Na and do not significantly contribute to the total loss.} The \edit1{difference in change of $\dot{M}_\mathrm{diff}$ and \edit1{thus $\dot{M}_\mathrm{ion}$} between species with continued magma ocean cooling} results from the \edit1{constantly dropping} total surface pressure \edit1{simultaneously to shifting partial pressures}. At low magma ocean surface temperatures, Na \edit1{exerts} most of the pressure, whereas at high temperatures condensing species, such as SiO, \edit1{are the predominant contributors} to the total pressure and mean molecular mass of the atmosphere (Fig.~\ref{fig:partvapP}). A lower surface pressure results in a lower homopause level and leads to a smaller diffusion area and rate\edit1{, however the rapid decline of the SiO partial pressure in the atmosphere ($n_\mathrm{SiO}/n_\mathrm{hom}$) results in the sharp drop of SiO homopause diffusion rates (i.e. Eq \ref{eq:diffusionrate}) but increases the $T_\mathrm{skin}$ and therefore the extent of the atmosphere (Fig. \ref{fig:hom_levels}). Similarly,} the loss rates through ionization at the exobase for a relatively low mean molecular weight (`light') \edit1{Na, K, and Fe based} atmosphere are up to a factor five \edit1{higher} than the $2\times10^5$~kg/s loss of Na from a 'heavy' atmosphere, which includes Mg and notably SiO.

These limits are significant for atmospheric escape estimates \edit1{by plasma heating, photoionization or Jeans escape. All three of those processes are} calculated from the exobase, \edit1{which is for plasma heating and photoionization }where ions \edit1{and photons} can access a rarefied neutral atmosphere. This is therefore diffusion-limited as a high flux is required to source the neutral \edit1{species} experiencing a momentum transfer from the plasma. Photoevaporation is not necessarily diffusion-limited so long as a sufficiently large column exists at the altitude where EUV photons are able to absorb on to infrared emitting molecules ($z_\mathrm{abs}$, Fig.~\ref{fig:mass_loss_schematic}). In \ce{N2}/\ce{CH4} atmospheres (e.g. Kuiper Belt Objects) the critical column density is estimated to be $\gtrsim\,$10$^{18}$/cm$^2$ \citep{Johnson2015} which is easily achieved at a fiducial absorption altitude of $z_\mathrm{abs}$ situated at $\sim$ 1.25 $R_P$, where the column density is equivalent to $\sim$10$^{21}$/cm$^2$ for an isothermal scale height of H $\approx$ 150~km at a $T_\mathrm{surf}$~=~2000~K. For a magma--silicate atmosphere, as studied here, SiO or a similar species would be able to re-emit in the infrared resulting in upper atmospheric expansion, and Roche-lobe overflow to space.

\subsubsection{Loss from volatile atmosphere}
In the volatile cases, \edit1{assuming a speciation as encountered at the H homopause}, the diffusion limited loss fluxes of primary species lie \edit1{between $10^4$~kg/s and $10^5$~kg/s} for all major species \edit1{(\ce{H2}, CO, \ce{H2O} and \ce{CO2})} in the small  and large proto-Mercury cases.

The loss flux of minor species Na, and K are several orders of magnitude lower \edit1{than those of the major species,} at $10^1$, and $10^0$~kg/s respectively. In the non-volatile cases, loss fluxes of Na and K are directly \edit1{proportional to their thermodynamic activities} in the melt. Sodium \edit1{activity} increases by about a factor 4.5 from the CB to NSP melt composition and K by a factor two between NSP source and EH4, respectively. Relative to non-volatile cases, loss fluxes for Na and K are several orders of magnitude lower in the high pressure, volatile-rich atmosphere at $T_\mathrm{surf}$~=~2000~K. 

If we assume the loss fluxes of the \edit1{dominant H, C, and O based species} at $T_\mathrm{surf}$~=~2000~K to be constant over the lifetime of the \edit1{molten surface} (Section~\ref{sec:res_cooling}) and integrate them for small and large Mercury volatile cases we obtain a total mass loss by photoionization of $4.1\times10^{16}$~kg and $1.8\times10^{17}$~kg, respectively. This exceeds the total photoionization mass loss from the \edit1{low absorbing N5} non-volatile case by only about one order of magnitude (Fig.~\ref{fig:BGintegrate}). The mass loss of Na \edit1{in the volatile cases}, however, only contributes about $10^{12}$~kg of the total, which is about four orders of magnitude below the total mass loss of the non-volatile cases. \edit1{Again, this assumes} that the Na loss flux is constant \edit1{in the volatile case. This} is deemed appropriate, because Na is only a minor component of such atmospheres, is lost at slow rates that represent an insignificant fraction of its total budget, and does not condense before reaching $T_\mathrm{surf}=1500$~K.


\section{Discussion} \label{sec:discussion}
\subsection{Mass loss of proto-Mercury}
\begin{figure*}
\gridline{
\fig{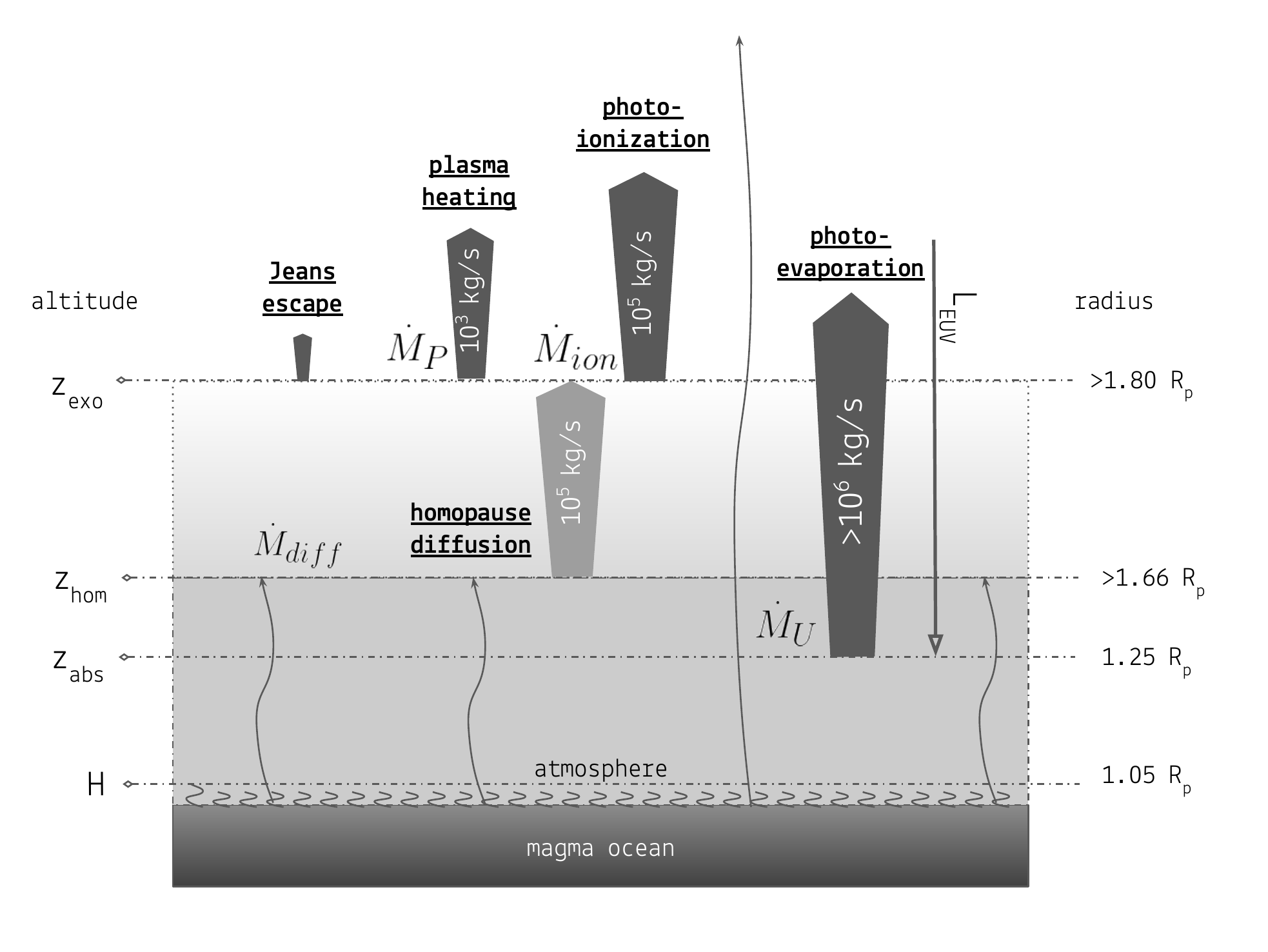}{0.99\textwidth}{}}
    \caption{Mass loss processes and their rates demonstrate the coupling between various atmospheric layers. $R_P$ is the planet radius, $z_\mathrm{exo}$ is the exobase altitude, and $z_\mathrm{hom}$ is the homopause altitude which governs exospheric loss processes of Jeans escape, plasma heating $\dot{M}_P$, and photo-ionization $\dot{M}_\mathrm{ion}$.  $z_\mathrm{abs}$ is the absorption altitude where upper-atmospheric heating (photoevaporation) $\dot{M}_U$ commences. The absorption altitude is assumed to lie below the homopause and therefore photoevaporation is not limited by homopause diffusion.
    }
    \label{fig:mass_loss_schematic}
\end{figure*}

Table \ref{tab:lossrates} tabulates the total atmospheric loss rates due to the following escape mechanisms: ionization $\Dot{M}_\mathrm{ion}$, photoevaporation $\dot{M}_U$, and plasma-heating $\dot{M}_P$. Figure \ref{fig:mass_loss_schematic} illustrates the atmospheric level from where the degassed magma ocean atmosphere is escaping. 
\begin{deluxetable}{@{\extracolsep{4pt}}c|ccc@{}}[!htbp]
\label{tab:lossrates}
\tablewidth{0pt} 
\tablecaption{Time-averaged mass loss rates in kg/s.}
\tablehead{
\multicolumn4c{Mass loss [log10]}   \\
\colhead{Process}    & \colhead{Size} &\multicolumn2c{Emissivity}\\
\cline{3-4} 
                 \colhead{}           &                          & \colhead{N5, N3}    & \colhead{V}
}
\startdata 
\cline{1-4}                                                                                             \\[-2ex]
\multirow{2}{*}{$\dot{M}_P$}                                            & S & 3.4           & 3.1            \\[1ex]
                                                                        & L & 3.4           & 3.2            \\[1ex]
\cline{1-4}                                                                                             \\[-2ex]
\multirow{2}{*}{$\dot{M}_\mathrm{ion}$ ($\dot{M}_{\mathrm{diff},i}$)}   & S & \edit1{5.6}  & \edit1{5.6}\\[1ex]
                                                                        & L & \edit1{5.2}  & \edit1{5.8}\\[1ex]
\cline{1-4}                                                                                                     \\[-2ex] 
\multirow{3}{*}{$\dot{M}_U$}                                            &   & \multicolumn2c{L$_\mathrm{EUV}$}  \\[1ex]
\cline{3-4}                                                                                                     \\[-2ex]
                                                                        &   & $t=1$~Myr      & $t=5$~Myr       \\[1ex]
 \multirow{2}{*}{$\eta_\mathrm{EUV}(10^{-3})$}                          & S & \edit1{7.5}   & \edit1{6.6}   \\[1ex]
                                                                        & L & \edit1{7.6}   & \edit1{6.7}   \\[1ex]
 \multirow{2}{*}{$\eta_\mathrm{EUV}(10^{-1})$}                              & S & \edit1{9.5}   & \edit1{8.6}   \\[1ex]
                                                                        & L & \edit1{9.6}   & \edit1{8.7}   \\
\enddata
\tablecomments{Loss rates of plasma heating $\dot{M}_P$, photoionization $\dot{M}_\mathrm{ion}$, and photoevaporation $\dot{M}_U$. Photoevaporation is insensitive to the atmosphere's emissivity and composition but depends on the EUV flux and the \edit1{EUV heating efficiency (end members of $10^{-3}$ and $10^{-1}$)}. The EUV flux is a function of the age of the solar system \citep[Eq. \ref{eq:lumin}, after][]{Johnstone2015,TCG15}.}
\end{deluxetable}


We find that the loss fluxes from the exobase caused by photoionization and atmospheric sputtering are supply-limited (in all volatile and non-volatile cases) by homopause diffusion $\dot{M}_\mathrm{diff}$, \edit1{which dictates the exobase elevation in order to remain in steady state}. The maximum $\dot{M}_\mathrm{ion}$ of volatile and non-volatile cases are comparable, even though their atmospheres are comprised of different major species \edit1{(non-volatile case: Na, and SiO, volatile case: \ce{H2}, \ce{H2O}, CO and \ce{CO2})}. This similarity is attributed to the higher $T_\mathrm{skin}$ of the non-volatile cases \edit1{at $T_\mathrm{surf}=2000$~K, caused by IR opacity which is tied to the mixing ratio of SiO. SiO pressure rapidly decreases with decreasing temperatures, which is contrary to \ce{CO_2} and \ce{H_2O} in the volatile cases. A higher skin temperature in the non-volatile cases hence compensates for the lower atmospheric pressures.} 

The photoevaporation rate is limited by the degree of upper atmospheric heating efficiency, $\eta_{EUV}$, rather than by $\dot{M}_\mathrm{diff}$ \edit1{, as well as by the supply of gases from surface evaporation (eq. \ref{eq:diff_surf}). The assumption of a constant photoevaporation rate is thus only valid for evaporation at high temperatures above 1600~K at which evaporation rates are fast enough for supply to be sustained, or for moderate mass loss rates of about 10$^{7}$~kg/s. The majority of mass loss occurs at high temperatures (Fig. \ref{fig:photoevap_integrated}) when surface evaporation rates are high compared to photoevaporative loss rates, the later of which are independent of temperature and depend instead on the EUV flux.}

The diffusion-limited loss \edit1{rates} by photoionization of the four major volatile species: \ce{H2}, \ce{H2O}, CO and \ce{CO2} from a volatile-rich atmosphere \edit1{total} $\sim 10^5$~kg/s. The loss of Na from a thick, volatile rich atmosphere is inhibited by its low mixing ratios at the homopause and exobase. Therefore, diffusion-limited loss of Na is most efficient when the atmosphere is thin, reaching a few $10^5$~kg/s, which coincides with \edit1{the total mass} loss rates from volatile cases. The total integrated mass loss by photoionization from Mercury's exosphere is low for \edit2{small} volatile and non-volatile cases with $\leq\,4.1\times10^{16}$ kg and $\leq\,3.0\times10^{16}$ kg, respectively.

The mass loss of \edit1{single species} is negligible compared to the total inventory of the magma ocean reservoir.  For example, 0.033~wt\% \ce{H2O} \edit1{and a low estimate of }0.1~wt\% Na in a total mass of M$_\mathrm{MO}\,\approx\,\times10^{23}$~kg \edit1{corresponds to a reduction of the total H$_2$O reservoir mass (volatile cases) and Na (non-volatile cases) by $\leq0.02\%$.} Assuming a well-mixed mantle reservoir, the bulk composition of Mercury would not significantly change even for species with low abundance in the reservoir and large loss rates such as \ce{H2O}, \ce{CO2}, and \ce{Na}. Energy-limited escape via photoevaporation, however, can erode up to \edit1{2.3~km} of Mercury's crust \edit2{which coincides with 0.3\% of small Mercury} (Figure~\ref{fig:photoevap_integrated}). \edit1{Assuming small heating efficiencies as well as a lower EUV flux, leads to integrated photoevaporation losses and eroded crust thicknesses that are reduced by four orders of magnitude.} 

Physical segregation between crystal and liquid during magma ocean cooling will induce chemical fractionation of element abundances with respect to those of the bulk mantle. Namely, the incompatible lithophile elements (Na, K, Al, and Ca) become enriched in late-stage liquids of a Hermean magma ocean. This effect is simulated by considering the composition of the NSP melt as a surface magma ocean analogue relative to that of its inferred source. These differences notwithstanding, the partial pressures of metal-bearing gas species vary only marginally among EH4, CB and NSP compositions. This is due to two factors: 1) vapor pressures of different elements vary by orders of magnitude among one another (e.g. compare Na with AlO) whereas abundances of these major elements vary only by a factor of 2--3 in most cases; and 2) higher mole fractions of Na and K in the NSP melt are partially compensated by their lower activity coefficients relative to the NSP source or EH4 composition. As we show in Section~\ref{sec:results}, all elements other than Na and K, and potentially Fe condense before reaching the exobase. We can therefore conclude that the atmospheric pressure and speciation around proto-Mercury only depends on the abundance of volatile- and moderately volatile elements.

\subsubsection{Early origin of surface Na}
In order to determine the potential impact of a magma ocean-generated atmosphere on the surface composition of a small proto-Mercury, we calculate the total mass of Na in the atmosphere. We consider a hypothetical scenario in which the atmosphere collapses as soon as the first crust forms, coinciding with the termination of the magma ocean stage when a surface temperature of 1500~K is reached. To obtain a result which is consistent with the notion of Na-poor building blocks \edit1{\citep[e.g.,][]{Humayun2000},} we use the low-Na CB composition (Table \ref{tab:mo_comp}) and its H-, and C-absent, pure Na atmosphere composition. The resulting total amount of Na integrated over the whole atmosphere yields about $10^{11}$~kg.

\edit1{Whether it precipitates as Na metal or as another compound depends on the composition of the atmosphere that exists. Although not considered in our model, such an atmosphere would contain significant quantities of other moderately volatile elements that could combine with Na to form complex molecules, namely, \edit1{F}, Cl and S. The species NaCl is inferred to be stable among volcanic gases \citep[e.g.,][]{Aiuppa2003} and has been directly observed in Io's atmosphere \citep{Lellouch2003, Moullet2010}, and may therefore be a potential candidate to form surficial deposits. Sodium chloride is also observed as a stable precipitate from experimentally-generated volcanic gas analogues \citep{Renggli2020}, and is therefore likely to occur as a Na-bearing phase on the Mercurian surface. This is supported by the coinciding distribution pattern of Na and Cl from more recent volcanic deposits found in gamma ray spectrometer data \citep[][]{Evans2015}.}

Here we consider \edit1{a simplified} case, for which the mass \edit1{of sodium in the atmosphere is uniformly distributed} over the surface of Mercury as pure, low density, Na metal, results in a layer less than 1~mm for the CB case. Using a more Na-rich composition like EH4 combined with an increased atmospheric reservoir size of large proto-Mercury would lead to a factor four thicker Na layer, but still less than 1 mm. In the volatile cases (i.e., with CO$_{2}$ and H$_{2}$O), the amount of Na in the atmosphere is identical to \edit1{the volatile-free cases, as, in our model,} the partial pressure of Na is independent of the presence of volatiles.\edit1{ The small dissolved quantities of CO$_{2}$ and H$_{2}$O in the silicate melt ($\leq$ 1000 ppm) should thereby not influence the activity coefficients of the major rock-forming species.} This hypothetical Na metal layer would not outlast meteorite impacts, which are assumed to have removed 50~m to 10~km of early crust \citep{Hyodo2021}. For the enrichment to be preserved, the atmospheric sodium would have to be incorporated into a layer with a thickness exceeding the removed crust. However, for a minimum layer of 50~m we obtain a total Na wt\% increase of merely $\sim$1~ppm and $\sim$10~ppb for small and large proto-Mercury cases, respectively. We thus conclude, that the collapse of an early Na-rich atmosphere would not contribute to a notable increase of Na in the surface.

\subsection{Controls on mass loss}

The mean column density at the exobase depends on the weighted average of the dominant species' collision cross sections (CCSs). Loss rates are directly related to the exobase density. However, using CCSs from \cite{Kim2002} that are about one order of magnitude smaller would reduce the homopause and exobase levels by a few tens of kilometers and decreases the homopause diffusion limited loss by $\leq2\%$. The sensitivity of mass loss to the chosen CCSs is therefore weak. 
In non-volatile cases, if we \edit1{consider} that all species except Na, K, and Fe condense \edit1{(FactSage in Section~\ref{sec:results}), then mean molecular mass and the collision cross section of the atmosphere decreases which enhances molecular diffusion (Eq. \ref{eq:moldiff}).  This pushes the homopause and therefore the exobase further from the planet surface, increasing the atmospheric surface area and therefore loss. Furthermore, the absence of SiO leads to a hotter skin temperature as the atmosphere becomes IR transparent, further enhancing loss. The difference of the ionization mass loss rate at the exobase between a Na, K, and Fe atmosphere and an atmosphere where SiO is a major component at high temperature is thereby about a factor three larger for all cases.}


\edit1{For rocky exoplanets on short orbits, the atmospheric temperature around our homopause levels ($10^{-7}$~bar) can be as high as 3800 K for a surface temperature of 2400 K \citep{Ito2015}. Mercury posesses different planet parameters (1~$M\mathrm{Earth}$ and 0.02~AU vs. 0.055~$M\mathrm{Earth}$ and 0.3~AU for Mercury), however, the more intense early UV flux experienced by Mercury could similarly boost the temperature at the homopause.  Calculations with an increased skin temperature of 3800~K at the homopause resulted in about a factor two higher photoionization loss rates for all cases.}

\edit1{We used photoevaporation as a proxy for thermally driven hydrodynamic escape. \cite{Krenn2021} has shown for a large range of parameters that photoevaporation can underestimate hydrodynamic escape especially at low EUV fluxes. Given our large incident EUV fluxes of about $10^2-10^3$~Js$^{-1}$m$^{-2}$ and our escape parameters (Table~\ref{tab:escape_parameters}) we expect to be within one order of magnitude of thermally-driven hydrodynamic escape rates \citep[compare EUV fluxes and escape parameters to Fig. 4 in][although our escape parameters for small Mercury lie just below the shown range]{Krenn2021}.}
\subsection{Atmospheric evolution and structure}
Figure \ref{fig:mass_loss_schematic} \edit1{illustrates the transport of mass away from different levels in the atmosphere}. In our model, atmospheric escape can be either energy-limited (e.g. $\dot{M}_U$) or diffusion-limited (Jeans escape, $\dot{M}_P$, $\dot{M}_\mathrm{ion}$). Below we describe the role of enhanced atmospheric heating or cooling on diffusion and energy-limited escape.

Diffusion rates are tied to the homopause density, which determines the homopause altitude. The eddy diffusion coefficient (\edit1{$K_{zz}$}) needed to determine $n_\mathrm{hom}$ bears large uncertainties, however. For all cases, a \edit1{$K_{zz}$} larger than the Earth-derived upper limit of $3.2\times10^6$ cm$^2$/s would most likely be adequate to accommodate proto-Mercury's increased atmospheric temperature, increasing $z_\mathrm{hom}$ and lowering $n_\mathrm{hom}$, leading to a slightly larger diffusion and therefore loss rate. Even if we assume a larger \edit1{$K_{zz}$}, however, homopause diffusion will remain the limiting factor for mass loss. We find for volatile cases, that even if \edit1{$K_{zz}$} is three orders of magnitude larger, \edit1{the total loss for volatile cases increases by a factor of less than two.} 
The sensitivity of $\dot{M}_\mathrm{diff}$ and therefore $\dot{M}_\mathrm{ion}$ to the eddy diffusion coefficient is therefore weak.

Ionization could further increase the exobase temperature, and hence the reported diffusion-limited loss fluxes could be a lower limit. \edit1{Whether mass loss occurs from the} exobase surface\edit1{, or whether it is the result of an advective outflow,} is canonically assessed by the escape parameter \edit1{\citep[e.g.][]{Genda2003}}. If the escape parameter $\lambda_0 \leq 3$ the atmosphere experiences \edit1{mass outflow due to its non-zero net velocity, and escape occurs inward of the exobase at the sonic point (where the thermal velocity exceeds the sound speed)}. If $\lambda_0 \gg 3$ the atmosphere escapes \edit1{because the mean free path is longer than the scale height, and Jeans escape prevails}.
Table \ref{tab:escape_parameters} shows how the exospheric escape parameters are all $4 \leq\lambda_0 \leq 15$, a \edit1{near-transitional} escape regime\edit1{ between Jeans- and hydrodynamic end members}, which was recently determined to be relevant for the putative magma ocean on the Moon \citep{tucker2021}. These authors demonstrated via DSMC simulations \citep{bird1994} that cooling due to escape is important for $\lambda_0 \lesssim 15$. Therefore in Table~\ref{tab:escape_parameters} based on our escape parameters, it would appear that although ionization may further enhance escape, cooling may temper this loss. In addition, the significant ionization rates of $\dot{M}_\mathrm{ion}\simeq$ 10$^6$~kg/s at the semi-major axis of proto-Mercury \edit1{promotes the generation of an ionosphere} that is modulated by the planetary magnetic field. Simulations on an early Mars analogue have demonstrated that ion escape is efficient at removing material \citep{egan2019}. Therefore, it is possible that we are underestimating escape by not considering magnetic interactions.

In the concurrent 'energy-limited' regime it would appear that if EUV photons are able to absorb on to a sufficiently high flux of molecules (Section~\ref{sec:nonvol_loss}), heating would overwhelm cooling. However, based on the escape parameters in Table \ref{tab:escape_parameters} it appears that cooling \edit1{associated with escape} may be important, arresting loss. At the same time, the study of low-mass, close-in exoplanets orbiting Sun-like stars has posited the idea that low-mass planets are nevertheless born with hydrogen/helium (H/He) envelopes although these are rapidly lost due to photoevaporation \citep{Mordasini2020}. For a H/He envelope equivalent to $1\%$ the mass of proto-Mercury, we find that our upper-limit on photoevaporation results in the dissipation of a H/He envelope in 10$^{4.4}$~years, which is larger than the \edit1{lifetime of the molten surface}. The possibility of a H/He envelope to persist during the \edit1{molten surface lifetime} is therefore non-trivial, and could result in significant heating which could not only enhance escape but also elongate the \edit1{melt} lifetime past the 10$^4$~years we study here.

\subsection{Origin and evolution of Mercury} \label{imp}

\edit2{The elevated core:mantle ratio, coupled with an Na- and S-rich surface, distinguish Mercury from the other terrestrial planets. Two key hypotheses exist to account for these characteristics; (1) the preferential loss of silicate material, either by evaporation \citep{Feg87} or by collisional stripping \citep{Benz1988} and (2) equilibrium condensation and sorting of metal from silicate in the solar nebula \citep{Lewis1972,Weidenschilling1978}.\\
In evaluating hypothesis (1), \cite{Feg87} concluded that $\sim$ 75--79\% of silicate material would need to be lost during a fractional vaporisation hypothesis to reproduce the core:mantle ratio of present-day Mercury. In this work, we show that such high fractions of loss of silicate material are untenable, be it from a small or a large proto-Mercury (total mass losses are below 0.3\%). The principal reason is that atmospheric cooling timescales are too rapid with respect to evaporation and escape timescales, meaning that integrated loss rates over $\sim$ 10$^{4}$ yr are small with respect to the mass of proto-Mercury. Moreover, substantial amounts of atmospheric- or collisional escape of Mercury's crust is not represented in the high K/U ratio of its surface \citep{Mccubbin2012}, as preferential loss of silicate material will predominantly deplete its incompatible lithophile element budget \citep{ONeill2008}.}

\edit1{There are several caveats to our conclusions, namely, that our results are valid for dry or C-, and H-bearing atmospheres, but do not consider the effect of other minor volatiles (Cl, S, F) on the volatility behaviour of metals. Metal chlorides and metal sulfides may be important gaseous species under moderate temperatures \citep[$\sim$ 1000 K,][]{Renggli2017}, increasing their volatility. Secondly, conditions on the surface of Mercury may have been considerably more reduced than modelled herein \citep[IW-5;][]{Cartier2019}. Because the partial pressures of most metal-bearing species increase with decreasing $f$O$_2$ (Eq. \ref{eq:evap1}) vaporisation rates for alkali metals may be an order of magnitude higher \citep[considering that their exponent, $n=1/4$;][]{Sossi2019}}.

\edit1{However, these faster evaporation rates may be offset by the presence of a surficial graphite layer on the magma ocean \citep{Keppler2019}. Such a layer is promoted under reducing conditions as the solubility of C in silicate melt decreases from $\sim$ 360 ppm at the IW buffer to 1 ppm at IW-4 \citep{Duncan2017,Keppler2019}. The extent of a graphite layer therefore depends on the C content of Mercury and its $f$O$_2$, both of which are poorly known.  A surficial lid would additionally delay cooling of the mantle, unless the lid is regularly broken as possibly occurred for the flotation crust on the moon \citep{PJE18}.  Nevertheless, the net effect of a graphite lid on Mercury's magma ocean would be to reduce the extent of degassing calculated herein.  Therefore, we conclude that the physico-chemical characteristics of Mercury cannot have been produced during a magma ocean stage on a near fully-grown planet.}

\edit1{These obstacles are ameliorated when considering vapour loss from planetary building blocks. Should Mercury have accreted from smaller, km-size planetesimals, then melting and vaporisation on the precursor bodies would have led to more efficient mass loss \citep[e.g.,][]{Hin2017}. Thus, vaporisation may still be a physically viable mechanism to explain Mercury's composition, provided it occurred on its precursor bodies. However, another problem arises because moderately volatile elements, such as Na, S and K, are always more volatile \edit1{(i.e., their partial pressures are higher for a given activity)} than the major mantle components, such as Mg and Si \citep{Sossi2019}. Moreover, as demonstrated herein, Na is more easily lost with respect to Mg and Si due to its lower molar mass and higher tendency to remain in the gas phase in an adiabatically-expanding atmosphere (\ref{sec:nonvol_loss}). As such, appealing to evaporative loss of Mg and Si to increase the core:mantle ratio while retaining Na and K is inconsistent with evaporation from a silicate melt on small planetary bodies. Therefore, other hypotheses should be considered.}

\section{Conclusions} \label{con}

We combined chemical and thermodynamic equilibrium models of the thermal evolution of Mercury's magma ocean and gaseous species derived thereof, to model the thermochemical evolution of an early atmosphere on Mercury.  For an initially large Hermean mantle with initial C and H budgets comparable to those of other rocky planets, namely Earth (`volatile cases'), the lifetime of \edit1{surficial melt} may have reached almost 10$^4$~years.  Compared to a present-day sized proto-Mercury without a greenhouse atmosphere, this lifetime is an order of magnitude larger and thereby may enable early atmospheric mass loss to occur over an extended duration.  Cases with C and H show that Mercury could have started with a 5--12~bar atmosphere. By contrast, excluding the presence of C and H species results in a thin, short-lived metal- and metal oxide-bearing atmosphere. The upper atmospheres of volatile cases are dominated by \ce{H2}, and CO whereas non-volatile cases are mostly Na and SiO.

Photoionization is a minor exospheric loss mechanism, limited by homopause diffusion ($\dot{M}_\mathrm{diff}$) up to a maximum of a few $10^5$~kg/s. If C and H volatiles are absent from the atmosphere, the $\dot{M}_\mathrm{diff}$ limit applies to SiO and Na.  Mass loss rates via photoevaporation, \edit1{$\dot{M}_U \leq$10$^{9.5}$~kg/s,} exceed those from all other known mechanisms due to the high EUV luminosity of the early Sun. This could \edit1{in the best case scenario} erode an equivalent thickness of up to \edit1{2.3 km} of proto-Mercury's crust when assuming high EUV heating efficiencies of $10^{-1}$.  Atmospheric sputtering $\dot{M}_U\sim$ 10$^{3.4}$~kg/s (also limited by $\dot{M}_\mathrm{diff}$) occurs at the exobase, knocking-off neutral gas molecules due to the ram pressure of the solar wind. 

By integrating atmospheric loss rates over \edit1{surficial melt lifetimes}, we bracket the expected total mass loss from Mercury's early atmosphere. Based on photoionization, Jeans escape, and plasma heating, the evaporation and loss of the magma ocean of proto-Mercury did not significantly modify its bulk composition.  This is because magma ocean cooling times are too short to drive substantial total loss for the determined atmospheric loss fluxes. Photoevaporation can remove an equivalent crustal thickness of \edit1{up to 2.3~km} in about 10,000 years, which is approximately $\sim$ 10$^{20}$ kg of material. Integrated losses of even the most volatile elements considered here, Na and K, are insignificant  with respect to their total budgets when escape is diffusion-limited \edit1{($\leq0.02\%$ decrease of the initial Na composition which would be a difference of $3\times10^{-4}$~wt\%)}.  Hence the present Na-rich surface composition may indicate that catastrophic volatile loss during the magma ocean stage did not occur, and that Mercury's peculiar composition is inherited from that of the solar-proximal region of the nebula from which it accreted.

\begin{acknowledgements}
Financial support has been provided by the Swiss National Science Foundation (SNSF) Fund (200021L182771/1).  D.~J.~Bower acknowledges SNSF Ambizione Grant 173992. P.~A.~ ~Sossi was supported by SNSF Ambizione grant 180025 and A.~Wolf by the National Science Foundation EAR 1725025 as well as the Turner Postdoctoral Fellowship. Thanks to S.~Suriano, P.~Saxena, A.~Heays, S.-M.~Tsai, and N~Ligterink for discussions relating to this work. Part of this work was conducted at the Jet Propulsion Laboratory, California Institute of Technology, under contract with NASA. 
\end{acknowledgements}

\software{SPIDER \citep{Bower2019,BSW18}, VapoRock \citep{Wolf2021}, VULCAN \citep{Tsai2017}, DISHOOM \citep{OZA19,gebekoza2020}, E-MC \citep{WULAM03}}

\clearpage

\appendix
\restartappendixnumbering


\section{Magma ocean model}
\label{app:magmaocean}
The evolving surface temperature of the Hermean magma ocean is calculated using the SPIDER code, which is described in detail in \cite{BSW18,BKW19,BHS21}.  Table~\ref{tab:mopara} shows the parameters used to model proto-Mercury.  The mass absorption coefficients of H and C volatile species are determined at 1.01 bar, and the coefficients of SiO at $3\times10^{-6}$ bar.

\begin{table}[!ht]
\centering
\caption{Standard parameters for magma ocean cases.}
\label{tab:mopara}
\begin{tabular}{lll}
\hline
\edit1{Parameter}	        & \edit1{Value}            & \edit1{Units}         \\
\hline
Core heat capacity 	        & 850 	                    & J kg$^{-1}$ K$^{-1}$	 \\
Core density			    & 7200 	                    & kg m$^{-3}$	         \\
Core radius			        & 2000                      & km	                 \\
Equilibrium temperature, $T_\infty$	    & 440 	                    & K                      \\
Gravity, $g$ 			    & Table~\ref{tab:mo_models}	& m s$^{-2}$             \\
Planetary radius, $R_P$	    & Table~\ref{tab:mo_models}	& km	                 \\
Boundary layer scaling, b   & $10^{-7}$                 & K$^{-2}$               \\
Al abundance$^{\dag\dag}$   & 19500                     & ppmw         \\
$^{26}$Al/Al (zero time) & $5.25\times10^{-5}$       & ---    \\
K abundance$^\dag$			& 403                       & ppmw	         \\
$^{40}$K/K (present time) & $1.17\times10^{-4}$         & ---    \\
Th abundance$^\dag$ 		& 49                        & ppbw	         \\
$^{232}$Th/Th (present time) & 1                         & ---    \\
U abundance$^\dag$			& 28                        & ppbw	         \\
$^{235}$U/U (present time)  & 0.007                     & ---        \\
$^{238}$U/U (present time)  & 0.993                     & ---        \\
\ce{H_2} mass absorption (CIA) & $5\times10^{-5}$          & m$^2$ kg$^{-1}$        \\
\ce{H_2} solubility law & \multicolumn{2}{l}{$^*$} \\
\ce{H_2O} mass absorption      & $10^{-2}$          & m$^2$ kg$^{-1}$        \\
\ce{H_2O} solubility law & \multicolumn{2}{l}{$^{**}$} \\
CO mass absorption          & $10^{-5}$          & m$^2$ kg$^{-1}$        \\
CO solubility law & \multicolumn{2}{l}{$^*$} \\
\ce{CO_2} mass absorption      & $10^{-4}$          & m$^2$ kg$^{-1}$        \\
\ce{CO_2} solubility law & \multicolumn{2}{l}{$^{**}$} \\
\ce{SiO} mass absorption (large)      & $10^{-3}$          & m$^2$ kg$^{-1}$        \\
\ce{SiO} mass absorption (small)      & $10^{-5}$          & m$^2$ kg$^{-1}$        \\
Initial surface temp        & 2400$^\ddag$              & K \\
\end{tabular}\\
\footnotesize{
$\dag\dag$ Average Al abundance based on the composition of EH4 and NSP source (Table~\ref{tab:mo_comp}).\\
$\dag$ Average current estimates for bulk heat source from \cite{TGP13} and natural abundances from \cite{RUE17}.\\ 
$\ddag$ Similar to maximum temperature estimate of Mercury's surface during accretion and differentiation \citep{BHS17}.\\
$*$ \cite{LBH21}.\\ 
${**}$ \cite{BKW19}.
}
\end{table}

\section{VapoRock species}
\label{app:oxides}
The species included in VapoRock are given in Table \ref{tab:oxides}.
\begin{deluxetable}{@{\extracolsep{10pt}}llllll@{}}[!ht]
\tablecaption{Species included in VapoRock \citep[][]{Wolf2021}.} \label{tab:oxides}
\tablehead
{
\multicolumn6c{Species}
}
\startdata
\ce{Al} & \ce{AlO}  & \ce{AlO2} & \ce{Al2}  & \ce{Al2O} & \ce{Al2O2} \\
\ce{Si} & \ce{SiO}  & \ce{SiO2} & \ce{Si2}  & \ce{Si2O2}& \ce{Si3} \\
\ce{K}  & \ce{KO}   & \ce{KO2}  & \ce{K2}   & \ce{K2O}  & \ce{     } \\
\ce{Na} & \ce{NaO}  & \ce{Na2}  & \ce{Na2O} & \ce{   }  & \ce{   } \\
\ce{Mg} & \ce{MgO}  & \ce{Mg2}  & \ce{   }  & \ce{   }  & \ce{   } \\
\ce{Ca} & \ce{CaO}  & \ce{Ca2}  & \ce{   }  & \ce{   }  & \ce{   } \\
\ce{Fe} & \ce{FeO}  & \ce{   }  & \ce{   }  & \ce{   }  &         \\
\ce{O}  & \ce{O2}    & \ce{   }  & \ce{   }  & \ce{   }  &         \\
\enddata
\end{deluxetable}{}

\section{Modified VULCAN}
\label{app:VULCAN}
We incorporated Na, Si, Mg, K, Fe and their derivatives into VULCAN by adding 12 reactions (from \url{kinetics.nist.gov}) to the preexisting chemistry network based on C, H and O (Table~\ref{tab:vulcreac}).  We initially added more reactions, but removed those that had a negligible impact on the resulting atmospheric speciation when omitted.


\begin{deluxetable}{@{\extracolsep{10pt}}lll@{}}[!ht]
\tablecaption{Key reactions added to VULCAN.} \label{tab:vulcreac}
\tablehead
{
\multicolumn3c{Reaction}
}
\startdata
\ce{OH + SiO }    & \ce{->} & \ce{SiO2 + H }  \\
\ce{OH + Si  }    & \ce{->} & \ce{SiO + H  }  \\
\ce{Si + O2  }    & \ce{->} & \ce{SiO + O  }   \\
\ce{NaO + O  }    & \ce{->} & \ce{Na + O2  }  \\
\ce{Na + H2O }    & \ce{->} & \ce{NaOH + H }  \\ 
\ce{H2O + NaO}    & \ce{->} & \ce{NaOH + OH}  \\
\ce{H2 + NaO }    & \ce{->} & \ce{NaOH + H }  \\
\ce{HCO + Na }    & \ce{->} & \ce{CO + NaH }\\
\ce{Mg + O2 }     & \ce{->} & \ce{MgO + O }   \\
\ce{H2O + KO}     & \ce{->} & \ce{KOH + OH}   \\
\ce{CO2 + Fe}     & \ce{->} & \ce{CO + FeO}   \\
\cline{1-3} 
\multicolumn3c{3-body reactions}\\
\cline{1-3}
\ce{OH + K + M}     & \ce{->} & \ce{KOH + M }\\
\ce{Na + O2 + M}  & \ce{->} & \ce{NaO2 + M}  \\
\ce{NaOH + M}  & \ce{->} & \ce{OH + Na + M}  \\
\ce{FeO + H2O + M}  & \ce{->} & \ce{Fe(OH)2 + M}\\
\enddata
\tablecomments{The reactions given affect the speciation of Si, Mg, Fe, Na and K and Si between $T$~=~2000 - 873 K and $P$~=~11.7 - 10$^{-7}$ bar.}
\end{deluxetable}{}


\section{Collision cross sections}
\label{app:crosssections}

The CCSs are shown in Table \ref{tab:crosssec}, which were approximated by the circular area of radius equal to the atom or bond length. Furthermore, all bonds were approximated to be covalent.
\begin{deluxetable}{lDllD}[!htbp]
\tablecaption{Species cross sections (CS) used} \label{tab:crosssec}
\tablehead
{
\colhead{Species} & \twocolhead{CS [$\text{\normalfont\AA}^2$]} &  & \colhead{Species} & \twocolhead{CS [$\text{\normalfont\AA}^2$]} 
}
\decimals
\startdata
H   & 0.88 &  & CO$_2$     & 10.3 \\
H$_2$  & 1.29 &  & KOH     & 19.6 \\
H$_2$O & 2.84 &  & Na      & 11.3 \\
O   & 1.13 &  & K       & 18.6 \\
O$_2$  & 4.08 &  & SiO     & 8.45 \\
C   & 1.41 &  & Mg      & 6.61 \\
CO  & 4.01 &  & Fe      & 7.65 
\enddata
\tablecomments{CS values are based on sizes of atomic, single, double, and triple bond data \citep[][]{Clementi1967,Pyykko2009b,Pyykko2009a,Pyykko2005}. }
\end{deluxetable}{}
\section{Atmospheric P--T profile}
\label{app:PTprofile}

\begin{figure}[tbhp]
\centering
\plotone{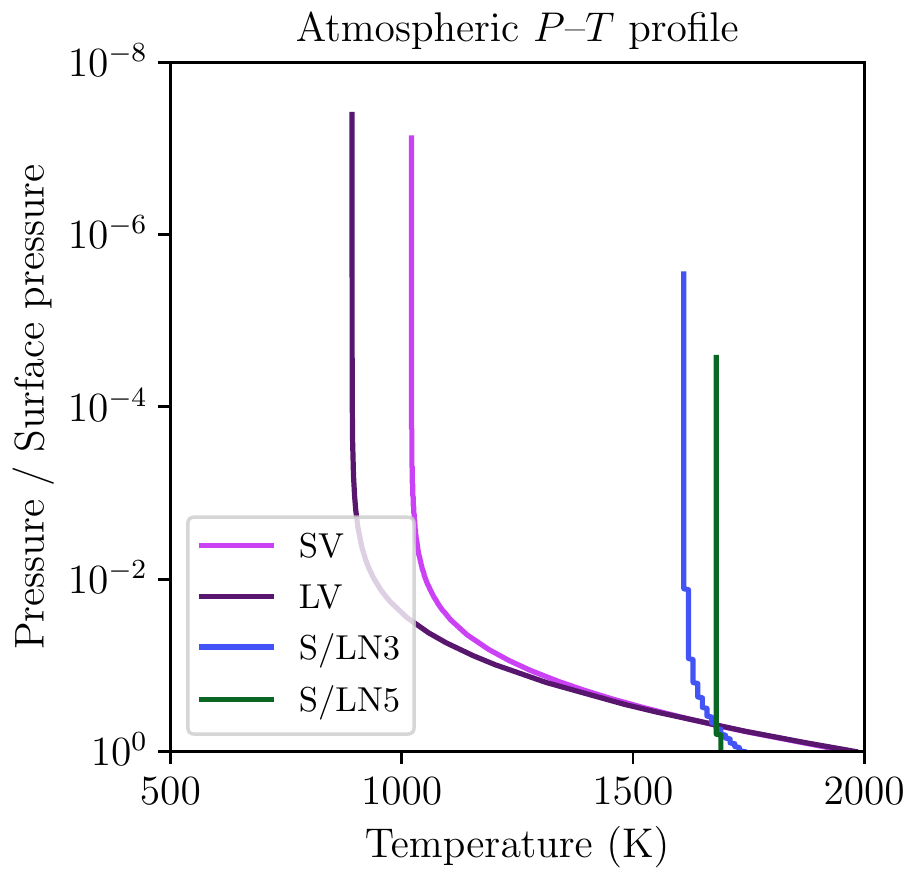}
\caption{\edit1{Atmospheric $P$--$T$ profiles of the volatile (V) and non-volatile (N3, N5) atmospheres surrounding a small (S) and large Mercury. The LN3 and LN5 profiles are omitted as they are visually indistinguishable from the respective S case profiles.}\label{fig:PTprofile}}
\end{figure}

SPIDER determines an atmospheric pressure--temperature profile through an analytical solution to the radiative transfer equations (see Appendix in \cite{AM85} and Sect.~3.7.2 in \cite{Andrews2010}).  The solution gives rise to the skin temperature equation (Eq.~\ref{eq:Tskin}).  Fig.~\ref{fig:PTprofile} shows the volatile (V) and non-volatile (N3, N5) atmospheric pressure--temperature profiles that are used for FactSage and VULCAN calculations.  An unphysical outcome of assuming only radiative equilibrium (no convection) is a temperature discontinuity between the base of the atmosphere and the surface of the magma ocean, which is visually more evident for the non-volatile cases that have a small optical depth.  Nevertheless, for all cases the surface temperature is 2000 K.

 \bibliographystyle{aasjournal}

\end{document}